\documentclass[12pt, onecolumn, draftcls,]{IEEEtran}

\usepackage[latin1]{inputenc} 
\usepackage{graphics, graphicx}
\usepackage{amssymb,amsmath}
\usepackage{algorithm}           
\usepackage{algpseudocode}   
\usepackage{cite}
\usepackage{multirow}
\usepackage{subcaption} 
\usepackage{fixmath}

\DeclareMathAlphabet\mathbfcal{OMS}{cmsy}{b}{n}
\newcommand{\eye}{\mathbf{I}}

\newcommand{\abs}[1]{\left\vert {#1} \right\vert}
\newcommand{\expect}[1]{\mathbb{E}\left[ {#1} \right]}

\newcommand{\vect}[1]{\mathbf{ {#1} }}

\newcommand{\rbrac}[1]{\left( {#1} \right)}
\newcommand{\sbrac}[1]{\left[ {#1} \right]}
\newcommand{\cbrac}[1]{\left\lbrace {#1} \right\rbrace}

\newcommand{\mm}{{\it massive} \textsc{mimo} }

\begin{document}


\title{Direction-of-Arrival Estimation Methods: A Performance-Complexity Tradeoff Perspective}

\author{Edno Gentilho Jr, Paulo Rogério Scalassara and Taufik Abrão\\ 
\normalsize Department of  Electrical Engineering (DEEL)
State University of Londrina  (UEL). \\
Po.Box 10.011, CEP:86057-970, Londrina, PR, Brazil. \\
\thanks{T. Abrao are with the Department of  Electrical Engineering;
State University of Londrina  (UEL). Po.Box 10.011, CEP:86057-970, Londrina, PR, Brazil.  Email: taufik@uel.br}
\thanks{Edno Gentilho Jr is with Federal Institute of Technology of Paraná (IFPR), Department of Electrical Engineering, Paranavaí­-PR. Email: edno.junior@ifpr.edu.br} 
\thanks{Paulo Rogerio Scalassara is with Federal Technological University of Paraná (UTFPR), Department of Electrical Engineering, Cornélio Procópio-PR. Email:prscalassara@utfpr.edu.br}     
}

\maketitle

\begin{abstract}
This work analyses the performance-complexity tradeoff for different direction of arrival (DoA) estimation techniques. Such tradeoff is investigated taking into account uniform linear array structures. Several DoA estimation techniques have been compared, namely the conventional Delay-and-Sum (DS), Minimum Variance Distortionless Response (MVDR), Multiple Signal Classifier (MUSIC) subspace, Estimation of Signal Parameters via Rotational Invariance Technique (ESPRIT), Unitary-ESPRIT and Fourier Transform method (FT-DoA). The analytical formulation of each estimation  technique as well the comparative numerical results are discussed focused on the estimation accuracy {\it versus} complexity tradeoff. The present study reveals the behavior of seven techniques, demonstrating promising ones for current and future location applications involving DoA estimation, especially for 5G \mm systems.
\end{abstract}


\section{Introduction } \label{introd}
\begin{figure}[!htbp]
\centering
\includegraphics[width=0.65\textwidth]{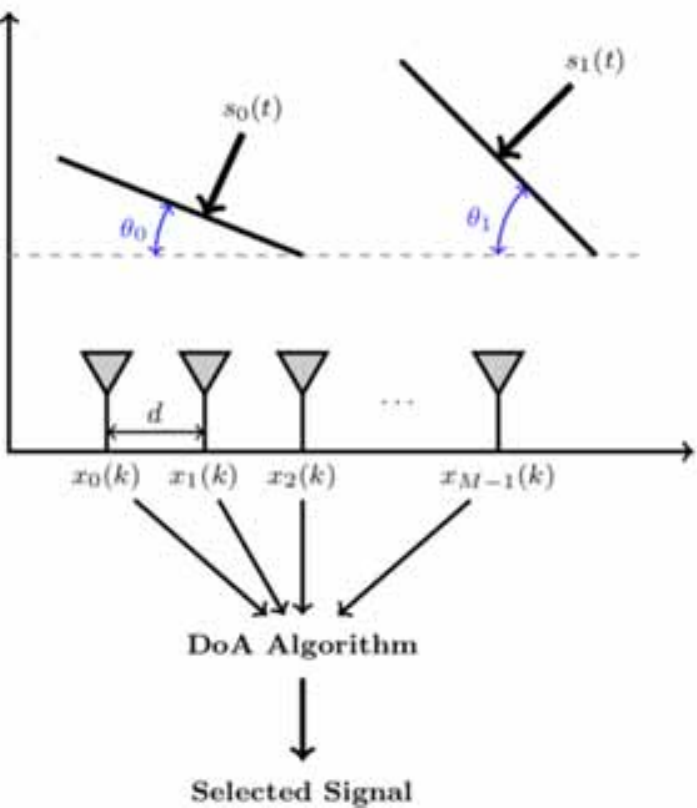}
\caption{Uniform linear array (ULA); $d$ is the distance  between the sensors; $\theta_i$ the elevation angle, and $M$ is the number of sensor array antennas.}
\label{fig:linear_array}
\end{figure}

Propagating fields are often measured by an array of sensors. A sensor array consists of a number of transducers or sensors arranged in a particular configuration. Each transducer converts an electromagnetic wave into a voltage. The electromagnetic waves are necessary for wireless communications systems implementation. Array signal processing applications include radar and wireless communication systems with  electromagnetic waves and sonar, seismic event prediction, microphone sensors with mechanical waves \cite{GODARA2004,monzingo,narrowband}. In a typical application, an incoming wave is detected by an array, the associated signals at different sensors in space can be processed to extract various types of information including their direction of arrival (DoA). The array model is illustrated in Fig. \ref{fig:linear_array}.
Moreover, the spatial-temporal estimation and filtering capability can be exploited for multiplexing co-channel users and rejecting harmful co-channel interference that may occur because of jamming or multipath effects. 

DoA algorithms can be divided into three categories: {\it extrema-searching techniques}\cite{capon69,reddy87,Schmidt1986,godara,haykin,rappaport,FGross2015}, {\it polynomial-rooting techniques}  \cite{Barabell1983} and {\it matrix-shift techniques} \cite{Roy1989,Haardt1997}. The matrix-shift techniques utilize estimates of the {\it signal subspace}  whereas most extrema-searching techniques and most polynomial-rooting techniques use estimates of its orthogonal complement, often referred to as {\it noise subspace}.

The DoA estimation process has been extensively researched since the 1980s, but the research area remains active, mainly due to recent and newly field of applications \cite{Balanis2007}, \cite{BenAllen2006}. Recent studies are mostly focused on specific applications and new approaches to the subject in order to improve performance while decrease computational complexity as well. For some DoA applications it is necessary to estimate the location of sources near and far from the array.  For instance, in \cite{two_stage2010} and \cite{two_stage2014} DoA techniques are discussed for far-field and near-field sources. To achieve greater accuracy in DoA, a large number of antennas are required, and this is not always feasible, either by physical space or cost limitations. 

An alternative is the {\it array virtualization} technique, which uses a small number of physical antennas, but able to estimate a much larger number of virtual antennas. The result is an estimation process with relative good precision combined to a reduced physical array dimensions, decreasing costs and physical space requirements \cite{nested2010}. However, with the virtualization of antenna arrays some problems arise, such as ambiguity in DoA estimated. An alternative to this disadvantage is the use of {\it coprime arrays}, which in addition to working with virtual antennas, is able to decrease or even eliminate the ambiguity in the DoA estimation \cite{cadis2014}, \cite{unfolded_coprime2017}. 

{Beamforming is used for directional signal transmission and reception with the versatility of changing amplitude and phase to help regulate power requirements and direct the beam to the desired direction. The bandwidth from 30 to 100 GHz, or millimeter wave (mmWave), is probably part of future mobile broadband as 5G communication systems are introduced into the global market. In the high-frequency transmission of mmWave, the significant loss of path during signal propagation limits the transmission range. To overcome this obstacle, directional antennas with beamforming capability are used for transmission and reception. The beamforming directs the antenna beams to the transmitter and receiver so that the data rate can be maximized with minimal loss.}

{In analog beamforming (ABF), a single signal is fed to each antenna element in the array by passing through analog phase-shifters where the signal is amplified and directed to the desired receiver. The amplitude/phase variation is applied to the analog signal at transmitting and where the signals from different antennas are added before the ADC conversion. Currently, analog beamforming is the most cost-effective way to build a beamforming array, but it can manage and generate only one signal beam. On the other hand, in digital beamforming (DBF), the RF signal at each antenna element is converted into two streams of binary baseband signals, $\cos(\cdot)$ and $\sin(\cdot)$, and used to recover both the amplitudes and phases of the signals received at each element of the array. The goal of this technology is the accurate translation of the analog signal into the digital realm. Matching receivers is a complex calibration process in which each antenna having its transceiver and data converters that generate multiple beams simultaneously from one array. The amplitude/phase variation is applied to the digital signal before DAC conversion at transmitting end. The received signals from antennas pass from ADC converters and DDC converters \cite{GODARA2004,mailloux2005,FGross2015}. Increasingly DBF techniques have being used more recently and works better with the challenges of the new 5G systems. This is the focus of application of the DoA methods analysed herein.}

Another very important point which has been widely studied by researchers is the computational complexity of DoA estimation methods.  With the new approaches on DoA, focused on the increasing capacity, there is a need for estimation methods and algorithms of low-complexity and high performance, aiming to be applied to real current and future systems \cite{lcdoa_mm2016, lcdoa_mm2016_2, lcdoa2016, unfolded_coprime2017, Zhang2011b}.

The {\it contribution} of this work consists in an extensive despite accurate comparison analysis involving different DoA estimation techniques; such analysis is carried out focusing on the  performance-complexity tradeoff under different performance metrics, focusing on massive-MIMO systems equipped with hundreds of antennas. Moreover, such tradeoff is investigated taking into account distinct DoA estimation methods, including:  a) Delay-and-Sum method (DS); b) Minimum Variance Distortionless Response (MVDR); c) Multiple Signal Classifier (MUSIC), d) Signal Parameters via Rotational Invariance Techniques (ESPRIT); e) Unitary-ESPRIT;  f) Fourier transform FT-DoA method; g) Root-MUSIC. In addition, the analysis is extended to large numbers of antennas, demonstrating the methods' viability under massive MIMO regime.

The remainder of this paper is divided as follows. Section \ref{sec:model} describes the model system deployed to formulate the DoA problem. An exploratory description of the current DoA techniques is given in  Section \ref{sec:DoA_Meth}. The complexity-performance comparative analysis for the DoA methods under different performance metrics is developed in Section \ref{sec:DoA_compare}. The main conclusions are offered in Section \ref{sec:concl}.

\section{Signal Model}\label{sec:model}
There are many topologies for the antenna arrays, such as basic linear and planar (UPA) ones, with sensors distributed uniformly or non-uniformly. With linear arrays it is possible to detect or estimate the direction (angle) of arrival of the signal in one dimension (1D), while with planar arrays it is possible to detect / estimate two dimensions or angles (2D); hence, with UPA it is possible to estimate the location of the source in elevation and azimuth. To simplify the analysis, this study deals with a linear arrangement only. In the following, the signal model is described taking into account the system geometry illustrated in Fig. \ref{fig:linear_array}.

Consider a uniform linear array (ULA) with $M$ sensors enumerated as $0,1,...,M-1$ and uniformly spaced into $d$ meters. At the array antenna elements,  the same far-field signal is steering in different time, and defined by: 
\begin{equation}
s(t)={\rm Re}\cbrac{s_\ell(t)e^{j2\pi f_ct}}
\end{equation} 
where $s_{\ell}$ is a $\ell$th narrowband source signal with the DoA to be estimated, {\it i.e.}, its bandwidth $B<<f_c$, where $f_c$ is the carrier frequency. Hence, the time-delay of arrival can be straightforward computed as:
\begin{equation}
\Delta t_k= \frac{kd}{c} \sin \theta,
\end{equation}
where $c= \lambda f_c$ is the velocity of propagation, $\lambda$ is the wavelength and $d$ is the regular distance between the antenna elements. Such distance must be $d\geq \frac{\lambda}{2}$ [m] to avoid ambiguity. Hence, the signal received by the $k$th antenna element is given by: 
\begin{equation}
x_k (t)= {\rm Re} \cbrac{s_\ell(t-\Delta t_k)e^{j2\pi f_c(t-\Delta t_k)}}
\end{equation}
Assuming that the received signal at the $k$th element is downconverted to the baseband, the baseband received signal is:
\begin{equation}
x_k (t)= s_\ell(t-\Delta t_k)e^{-j2\pi f_c\Delta t_k}
\end{equation}
The received baseband signal is sampled with sampling period $T$ seconds, which is also the symbol period
\begin{equation}
x_k (nT)= s_\ell(nT-\Delta t_k)e^{-j2\pi f_c\Delta t_k}.
\end{equation}

In a wireless digital communication system, the symbol period $T$ is much greater than each of the propagation delay across the array elements:
\begin{equation}
T>>\Delta t_k,\qquad k=0,1,...,M-1. 
\end{equation}
This allows the following approximation to be made \cite{godara}:
\begin{equation}
\label{eq:approximation}
x_k (nT) \approx s_\ell(nT)e^{-j2\pi f_c\Delta t_k}.
\end{equation}

A discrete-time notation  with time index $n$ is now introduced; hence, eq. (\ref{eq:approximation}) can be re-written as:
\begin{equation}
x_k \sbrac{n} \approx s_\ell\sbrac{n}e^{-j\frac{2\pi}{\lambda}kd \sin \theta}= s_\ell\sbrac{n}a_k(\theta)
\end{equation}
where $a_k(\theta) = e^{-j\frac{2\pi}{\lambda}kd \sin \theta},$\,\, for $k=0,1,...,M-1$. 

Supposing that there are $\mathcal{L}$ DoA far-signals to be estimated, the $n$th symbol of the $\ell$th signal is denoted by $s_\ell\sbrac{n}$ for $\ell = 0, 1,..., \mathcal{L}-1$. Then, the baseband, sampled signal at the $k$th antenna element can be expressed as:
\begin{equation}
\label{eq:signal}
x_k\sbrac{n} \approx \sum_{\ell=0}^{\mathcal{L}-1} s_\ell\sbrac{n}a(\theta_\ell).
\end{equation}

\subsection{Matrix Representation for Array Data}
Considering the array antenna elements $k=0,1,2,...,M-1$, eq. \eqref{eq:signal} can be re-written in a matrix form:
\begin{equation}\label{eq:array_data_vector}
{\bf x}_n=\sbrac{{\bf a}(\theta_0)\,{\bf a}(\theta_1)\, ...\, {\bf a}(\theta_{\mathcal{L}-1})}{\bf s}_n+{\bf n}_n = {\bf A}{\bf s}_n+{\bf n}_n
\end{equation}
where ${\bf x}_n$ is the $M$-dimensional sampled signal vector, ${\bf A}$ is the $M \times \mathcal{L}$ array matrix, ${\bf s}_n$ is the received signal vector and  $n_k[n]$ is the additive noise considered at each element. 
Notice that matrix ${\bf A}$ is formed by the column-vectors ${\bf a}(\theta_\ell)$, namely vectors of direction of the signals $s_\ell(t)$, defined by:
\begin{equation}
\label{eq:a_matrix}
{\bf A}
=
\left[ {\begin{array}{*{20}{c}}
{\bf a}(\theta_0)& \cdots &{\bf a}(\theta_\ell)&\cdots&{\bf a}(\theta_{\mathcal{L}-1})\\
\end{array}} \right],
\end{equation}
\begin{equation}
\label{eq:atheta_matrix}
\text{with} \qquad {\bf a}(\theta_\ell)
=
\left[ {\begin{array}{*{20}{c}}
a_0(\theta_\ell)\\
 \vdots  \\
a_{M-1}(\theta_\ell)
\end{array}} \right].
\end{equation}

Assuming that the DoA of $\mathcal{L}$ signals are different, then the vectors form a linearly independent set. The vector ${\bf n}_n$ represents the uncorrelated thermal noise aggregated to the $M$ array antenna elements. Since the direction-vectors are a function of the DoA of $\mathcal{L}$ source-signals, the angles can be calculated if the direction vectors are known or if a basis for the subspace spanned by these vector can be determined.

\subsection{Eigenstructure of the Spatial Covariance Matrix}
Considering vectors ${\bf s}_n$ and ${\bf n}_n$ uncorrelated and ${\bf n}_n$ additive  white Gaussian noise samples with zero mean and covariance matrix $\sigma^2 \eye$. Defining the spatial correlation matrix as: 
\begin{equation}
\begin{array}{ccl}
{\bf R}& = & \expect{{\bf x}_n {\bf x}_n^H}\\
& = & \expect{({\bf As}_n + {\bf n}_n)({\bf As}_n + {\bf n}_n)^H}\\
& = & {\bf A} \expect{{\bf s}_n {\bf s}_n^H} {\bf A}^H + \expect{{\bf n}_n {\bf n}_n^H}\\
& = & {\bf A}{\bf R}_{\rm ss}{\bf A}^H + \sigma^2 \eye_{M \times M}.
\end{array}
\end{equation}
Since ${\bf R}$ is Hermitian, it can be unitarily decomposed with real eigenvalues. Let us examine the eigenvectors of the spatial correlation matrix ${\bf R}$ and assume that $M$ is large enough, i.e., $M>\mathcal{L}$. Any vector, ${\bf q}_n$, which is orthogonal to the columns of ${\bf A}$, is also an eigenvector of ${\bf R}$, which can be shown by manipulating the characteristic equation:
\begin{equation}
\vect{R}\vect{q}_n = ({\bf A}{\bf R}_{\rm ss}{\bf A}^H + \sigma^2 \eye)\vect{q}_n = {\bf 0}+\sigma^2 \eye \vect{q}_n = \sigma^2 \vect{q}_n.
\end{equation}

The corresponding eigenvalue of $\vect{q}_n$ is equal to $\sigma^2$. Because $\vect{A}$ has dimension $M \times \mathcal{L}$, there will be $M-\mathcal{L}$ linearly independent vectors whose eigenvalues are equal to $\sigma^2$. The space spanned by the $M-\mathcal{L}$ eigenvectors is called the {\it noise subspace}. If $\vect{q}_s$ is an eigenvector of $\vect{AR}\vect{A}^{H}$ then,
\begin{equation}
\vect{R}\vect{q}_s = ({\bf A}{\bf R}_{\rm ss}{\bf A}^H + \sigma^2 \eye)\vect{q}_s = \sigma_s^2 \vect{q}_s+\sigma^2 \eye \vect{q}_s = (\sigma_s^2 + \sigma^2) \vect{q}_s
\end{equation}  
Notice that $\vect{q}_s$ is also an eigenvector of $\vect{R}$ with eingenvalue $(\sigma_s^2 + \sigma^2)$, where $\sigma_s^2$ is the eigenvalue of ${\bf A}{\bf R}_{\rm ss}{\bf A}^H$. Since ${\bf A}{\bf R}_{\rm ss}{\bf A}^H \vect{q}_s$ is a linear combination of columns of $\vect{A}$, the eigenvector $\vect{q}_s$ lies in the column-space of $\vect{A}$. These are $\mathcal{L}$ such linearly independent eigenvectors of $\vect{R}$. Then, the space spanned by this $\mathcal{L}$ vectors is the {\it signal subspace}. The signal and noise subspaces are orthogonal each other.

Finally, the eigen-decomposition of $\vect{R}$ can be written as
\begin{equation}
\label{eq:eigendecomposition}
\vect{R}=\vect{QDQ}^H
=
\left[ {\begin{array}{*{20}{c}}
\vect{Q}_s & \vect{Q}_n 
\end{array}} \right]
\left[ {\begin{array}{*{20}{c}}
\vect{D}_s & {\bf 0}\\
{\bf 0} & \sigma^2 \eye
\end{array}} \right] 
\left[ {\begin{array}{*{20}{c}}
\vect{Q}_s & \vect{Q}_n 
\end{array}} \right]^H
\end{equation}

The matrix $\vect{Q}$ is partitioned into an $M \times \mathcal{L}$ matrix $\vect{Q}_s$ whose columns are composed by the $\mathcal{L}$ eigenvectors corresponding to the signal subspace, and an $M \times (M-\mathcal{L})$ matrix $\vect{Q}_n$ whose columns correspond to the noise eigenvectors. The matrix $\vect{D}$ is a diagonal matrix whose diagonal elements are the eigenvalues of $\vect{R}$ and is partitioned into an $\mathcal{L} \times \mathcal{L}$ diagonal matrix $\vect{D}_s$ whose diagonal elements are the signal eigenvalues and an $(M-\mathcal{L})\times(M-\mathcal{L})$ scaled identity matrix $\sigma^2\eye_{M \times M}$ whose diagonal are composed by $M - \mathcal{L}$ noise eigenvalues.

\section{DoA Estimation Methods}\label{sec:DoA_Meth}

\subsection{Extrema-Searching Techniques}
Extrema-searching techniques work making a beam scan in the spacial dimension while measuring the received power by the array sensors. The highest power peaks are the DoA possible estimates.

In this section, Delay-and-Sum method (DS) \cite{GODARA2004}, the Minimum Variance Distortionless Response (MVDR) method \cite{haykin} and Multiple Signal Classifier (MUSIC) method \cite{Schmidt1986} will be revisited. DS and MVDR are essentially based on beamforming, while MUSIC utilizes noise subspace, resulting in a high-resolution estimation. 
The discourse of this section commences with the basic ULA model, where the signal ${y}(n)$ is given simply by the weighted sum of the signal received by the array sensors:
\begin{equation}
\label{eq:conventional}
y(n)={\bf w}^H{\bf x}(n).
\end{equation}
%
\subsubsection{Delay-and-Sum Method}
The DS method calculates the DoA by measuring the signal strength at each possible arrival angle (scanning) and selecting the arrival angles at  power peaks\cite{godara}.  In the case of weights ${\bf w}$, according to (\ref{eq:conventional}), equal to the steering vector, it will occur a power peak in the beam. The highest power point corresponds to the estimated angle of arrival. The output mean power of the beamformer using this method is given by:
\begin{equation}
\begin{array}{ccl}
\vspace{2mm}
P_{\textsc{ds}}(\theta) & = & \expect{\abs{y(n)}^2}\\
\vspace{2mm}
& = & {\bf w}^H \expect{{\bf x}(n){\bf x}^H(n)}{\bf w}\\
& = & {\bf w}^H{\bf R}{\bf w}
\end{array}
\end{equation} 

Let $s(n)$ arriving with steering angle $\theta_0 $. Based on the model in (\ref {eq:signal}) the average received power can be defined as:
\begin{equation}
\begin{array}{ccl}
\vspace{2mm}
P_{\textsc{ds}}(\theta_0) & = & \expect{\abs{{\bf w}^H{\bf x}(n)}^2}\\
\vspace{2mm}
&  = &\expect{\abs{{\bf w}^H\left[{\mathbf a}(\theta_0){\mathbf s}(k)+{\mathbf n}(n)\right]}^2}\\
& = & \rbrac{\abs{{\mathbf w}^H{\mathbf a}(\theta_0)}^2 \rbrac{\sigma^2_s + \sigma^2_n}} \\
\end{array}
\end{equation} 
where ${\mathbf a}(\theta_0)$ is the direction vector associated with the angle $\theta_0$, ${\mathbf n}(n)$ is the noise vector, $\sigma^2_s$ and $\sigma^2_n$ is the signal power and noise power respectively. The average received power intensity has its maximum value when ${\mathbf w}= {\mathbf a}(\theta_0)$. So, of all the possible weight vectors, the receiving antenna will have the biggest gain in the direction $\theta_0$, when ${\mathbf w}= {\mathbf a}(\theta_0)$. This is because ${\mathbf w}= {\mathbf a}(\theta_0)$ aligns the phases of the components of arrival signal of $\theta_0$ in the sensors. In \textsc{ds} method, a scan is performed on all possible angles of arrival and the power measurement is performed on all of them. The mean power of steering angle is: 
\begin{equation}
\begin{array}{ccl}
\vspace{2mm}
P_{\textsc{ds}}(\theta) & = & {\bf w}^H{\bf R}{\bf w}\\
\vspace{2mm}
&  = & {\mathbf a}^H(\theta){\mathbf R}{\mathbf a}(\theta)
\end{array}
\end{equation}
Hence, the arrival angles $ \theta $ are determined by evaluating the power peaks.

Despite being computationally simpler, the width and height of the side lobes limit the performance (discrimination capability) and effectiveness of the DS method when signals from multiple directions / sources are involved, implying in poor resolution. One way to improve it consists of increasing the number of sensors, thus increasing the elements of vector $ {\mathbf a} (\theta)$, which increases the delay-sum signal processing and complexity. 

The pseudocode of the DS-DoA method is depicted in Algorithm \ref{ds code}, where $M$ is the number of antennas, $\mathcal{L}$ is the number of sources, $S$ the number of samples and $P$ is the number of scan steps of $\theta \in \{-90^{\circ}:90^{\circ}\}$. The complexity analysis is based on \cite{trench1989, complex2007, cybenko1980}. 
\begin{algorithm}
\caption{Delay \& Sum (DS-DoA) Procedure} \label{ds code}
\begin{algorithmic}[1]
\State {$\vect{\hat R} =\frac{1}{S}\sum_{n=0}^{S-1}{\bf x}_n {\bf x}_n^H$} \Comment{Autocor. Matrix - {$SM^2+M^2$}}
\vspace{1mm}
\State $P_{\textsc{ds}}(\theta)={\mathbf a}^H(\theta){\mathbf R}{\mathbf a}(\theta)$ \Comment{Angles scan step - {$M^2+M$}}
\vspace{1mm}
\State Findpeaks\Comment{Determine estimated DoA - $4\mathcal{L}P$}
 \end{algorithmic}
 \Comment{Total complexity: $M^2(S+2)+M+4\mathcal{L}P$}
\end{algorithm}
%
\subsubsection{MVDR Method}
Capon's minimum variance distortionless response or MVDR is similar to the delay-and-sum technique, since it evaluates the power of the received signal in all possible directions. The power from the DoA with angle $\theta$ is measured by constraining the beamformer gain to be 1 in that direction and using the remaining degrees of freedom to minimize the output power contributions of signals coming from all other directions. The optimization problem can be stated mathematically as a constrained minimization problem. The idea is that for each possible angle $\theta$, the power in the cost function must be minimized w.r.t. ${\bf w}$ subject to a single constraint: 
\begin{eqnarray}
\label{eq:mvdr}
\min_{\bf w} & {{\bf w}^H{\bf Rw}}\notag\\
\text{s.t.} & \,\, {\bf w}^H{\bf a}(\theta) = 1
\end{eqnarray}
resulting in the MVDR received power solution:
\begin{equation}
P_\textsc{mvdr}(\theta)=\frac{1}{{\mathbf a}^H(\theta){\mathbf R}^{-1}{\mathbf a}(\theta)}
\end{equation}

The disadvantage of this method is that an inverse matrix computation is required which may become poor conditioned if highly correlated signals are present. This method, however, provides higher resolution than the delay-and-sum method.  A pseudocode for the MVDR-DoA is provided in Algorithm \ref{mvrd code}.
\begin{algorithm}[H]
\caption{MVDR-DoA  Procedure }\label{mvrd code}
\begin{algorithmic}[1]
\State {${\vect{\hat R}} =\frac{1}{S}\sum_{n=0}^{S-1}{\bf x}_n {\bf x}_n^H$} \Comment{Autocorr. Matrix - {$SM^2+M^2$}}
\vspace{1mm}
\State ${\vect{\hat R}}^{-1}$ \Comment{Matrix inverse - {$4M^2$}}
\vspace{1mm}
\State $P_{\textsc{mvdr}}(\theta)=\frac{1}{{\mathbf a}^H(\theta){\vect{\hat R}}_{\rm xx}^{-1}{\mathbf a}(\theta)}$ \Comment{Angles scan - {$M^2+M$}}
\vspace{1mm}
\State Findpeaks\Comment{Determine estimated DoA - $4\mathcal{L}P$}
\end{algorithmic}
\Comment{Total complexity: {$M^2(S+6)+M+4\mathcal{L}P$}}
\end{algorithm}

\subsubsection{Multiple Signal Classifier (MUSIC) DoA Estimator}
The steering vectors corresponding to the incoming signals lie in the signal subspace; therefore, they are orthogonal to the noise subspace. One way to estimate the DoAs of multiple signal source is to search through the set of all possible steering vectors and  find those that are orthogonal to the noise subspace. \textsc{music} DoA estimator implements such strategy. If ${\bf a}(\theta)$ is the steering vector corresponding to one of the incoming signals, then ${\bf a}(\theta)^H{\bf Q}_n = 0$, where ${\bf Q}_n$ is the noise subspace matrix. In practice, ${\bf a}(\theta)$ will not be precisely orthogonal to the noise subspace due to errors in estimating ${\bf Q}_n$. However the function
\begin{equation}
\label{eq:musicequation}
{P_\textsc{music}}(\theta) = \frac{1}{{\bf a}^{H}(\theta){\bf Q}_{n}{\bf Q}_{n}^{H}{\bf a}(\theta)}
\end{equation} 
implies a very large value when $\theta$ is equal to the DoA related to one of the signals. ${P_\textsc{music}}(\theta)$ function is known as a pseudo "spectrum" provided by MUSIC.

In terms of implementation, the MUSIC-DoA first estimates a basis for the noise subspace, ${\bf Q}_n$, and then determines the  $\mathcal{L}$ peaks in (\ref{eq:musicequation}); the associated angles provide the DoA estimates. A pseudocode for the MUSIC-DoA procedure is described in the Algorithm \ref{music code}.
\begin{algorithm}
\caption{MUSIC-DoA Procedure}\label{music code}
\begin{algorithmic}[1]
\State {$\vect{\hat R} =\frac{1}{S}\sum_{n=0}^{S-1}{\bf x}_n {\bf x}_n^H$} \Comment{Autocorr. Matrix - {$SM^2+M^2$}}
\vspace{1mm}
\State $\vect{\hat R}=\vect{QDQ}^H$ \Comment{Eigendecomposition - {$\frac{2M^3}{3}$}}
\vspace{1mm}
\State {${\bf Q}_{n}{\bf Q}_{n}^{H}$} \Comment{Eigenvectors multiplication - {$M^3+M^2\mathcal{L}$}}
\vspace{1mm}
\State $ P_\textsc{music}(\theta) = \frac{1}{{\bf a}^{H}(\theta){\bf Q}_{n}{\bf Q}_{n}^{H}{\bf a}(\theta)}$ \Comment{Angles scan- {$M^2M$}}
\vspace{1mm}
\State Findpeaks\Comment{Determine estimated DoA - $4\mathcal{L}P$}
\end{algorithmic}
\Comment{Total complexity: $\frac{5}{3}M^3+M^2(S+1+\mathcal{L}+M)+4\mathcal{L}P$}
\end{algorithm}

\subsection{Matrix-Shifting Techniques}
In this subsection, matrix-shifting based techniques are revisited, more specifically Estimation of Signal Parameters via Rotational Invariance Techniques (ESPRIT), which is one of the most widely used method for DoA estimation. As previously mentioned, the MUSIC method uses the noise subspace while ESPRIT deploys the signal subspace in conjunction with a rotational variance technique. 

The ESPRIT method was first introduced in \cite{Roy1989}; the ESPRIT-based DoA estimates are obtained neither nonlinear optimization nor search of any spectral measure; hence, it results in a computational complexity lower than the extrema-searching methods, scanning for all possible angles of arrival. 

\subsubsection{Conventional ESPRIT Method}
\begin{figure}[!htbp]
\centering
\includegraphics[width=0.70\textwidth]{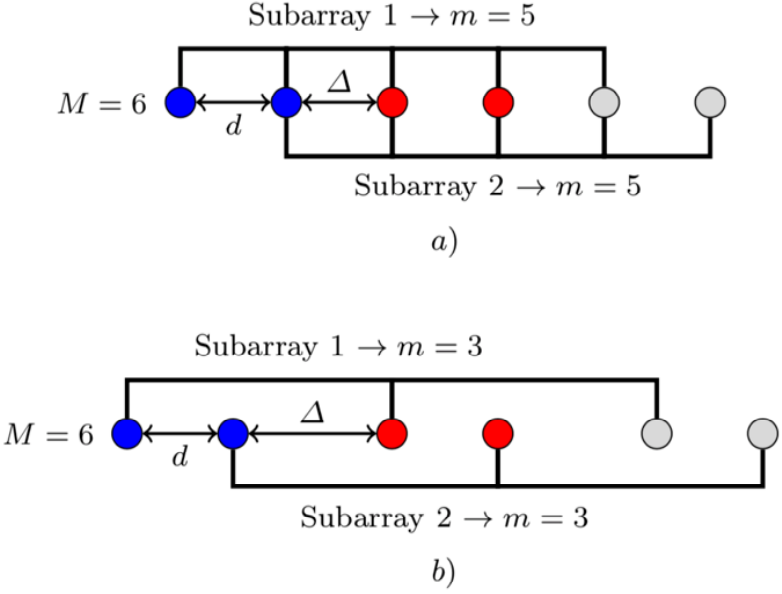}
\caption{Two examples of ESPRIT subarrays formation using $M=6$ antenna elements: {\bf a}) equidistant array with 3 equidistant identical doublets and  $d = \Delta$. {\bf b}) array with 3 non-equidistant identical doublets and  $d \neq \Delta$. } 
\label{fig:array_esprit}
\end{figure}

The ESPRIT operates under an array of antennas with $M$ elements, divided into sensor doublets as shown in Fig. \ref{fig:array_esprit}. Each sensor is distant $d$ from its respective pair and each doublet is distant $\Delta$ from one another. The doublets can be separated to form 2 subarrays with $m$ elements in each. The distance $d$ may be different from $\Delta$ as shown in Fig. \ref{fig:array_esprit}b, which makes it quite dynamic in cases of non-uniform arrays. However, the most commonly used antenna arrays possess sensors uniformly spaced, as depicted in Fig. \ref{fig:array_esprit}a. Then in this work the uniform array configuration has been adopted. 

The subarrays are represented by ${\bf x}_{1}$ and ${\bf x}_{2}$. The output of the ${\bf x}_1$ and ${\bf x}_2$ subarrays is expressed as \cite{Roy1989, Haardt1997}:

\begin{subequations}
\begin{align}
{\bf x}_1[n] & =  \sum^{\mathcal{L}-1}_{\ell = 0} s_\ell[n]{\bf a}(\theta_\ell)+ {{\bf n}_{\rm x_1}[n]},\\
{\bf x}_2[n] &=  \sum^{\mathcal{L}-1}_{\ell = 0} s_\ell[n]e^{j\frac{2\pi}{\lambda} \Delta \sin(\theta_\ell)} {\bf a}(\theta_\ell)+ {{\bf n}_{\rm x_2}[n]},
\end{align}
\end{subequations}
for $n=1, 2, \ldots, S$ samples; besides, ${\bf x}_1$ and ${\bf x}_2$ are $m \times 1$ vectors, ${\bf n}_{\rm x_1}$ and ${\bf n}_{\rm x_2}$ are the $m \times 1$ vectors representing the noise samples at the input of two subarrays, respectively. Writing in matrix form, the output of the subarrays ${\bf x}_1$ and  ${\bf x}_2$ can be expressed as  \cite{Roy1989, Haardt1997}:
\begin{subequations}
\begin{align}
{\bf x}_1&= {\bf As}+ {\bf n}_{\rm x_1} \label{eq:x1_matrix}\\
{\bf x}_2&= {\bf A \Phi s} + {\bf n}_{\rm x_2}. \label{eq:x2_matrix}
\end{align}
\end{subequations}
where ${\bf \Phi} = {\rm diag}\cbrac{e^{j\frac{2\pi}{\lambda} \Delta \sin \theta_0}, ... ,e^{j\frac{2\pi}{\lambda} \Delta \sin \theta_{\mathcal{L}-1}}} $ is a $\mathcal{L} \times \mathcal{L}$ diagonal matrix relating the signals received by the two subarrays, named the rotational operator \cite{Roy1989}. Notice that matrix $\bf \Phi$ in \eqref{eq:x2_matrix} represents an extra delay caused by ${\Delta}$ on the second subarray ${\bf x}_2$. Combining Eq. (\ref{eq:x1_matrix}) and (\ref{eq:x2_matrix}) the vector of total array output is formed \cite{Roy1989, Haardt1997}: 
\begin{equation}
{\bf x}[n] = 
\begin{bmatrix}
    {\bf x}_1[n] \\
    {\bf x}_2[n] \\
\end{bmatrix}
=\begin{bmatrix}
    {\bf A} \\
    {\bf A \Phi} \\
\end{bmatrix}
{\bf s}[n] +
\begin{bmatrix}
    {\bf n}_1[n] \\
    {\bf n}_2[n] \\
\end{bmatrix}
= {\bf Q}_s{\bf s}[n] + {\bf n}[n].
\end{equation}

The ${\bf Q}_s$ structure is exploited to estimate the diagonal elements of ${\bf \Phi}$ without knowing {\bf A}. The ${\bf Q}_s$ columns {span} the signal subspace of the concatenated subarrays. Hence,  ${\bf Q} = \left[ {\begin{array}{*{20}{c}}
{\bf Q}_s & {{\bf Q}_n}\\\end{array}} \right]$ is obtained by the eigen-decomposition of $\bf R$ from Eq. (\ref{eq:eigendecomposition}). If ${\bf E}_s$ is a matrix whose columns form a basis for the subspace of signal corresponding to the data vector ${\bf x}$, then ${\bf Q}_s$  and ${\bf E}_s$ are related by a $\mathcal{L} \times \mathcal{L}$ transformation ${\bf T}$ \cite{Roy1989} expressed by:
\begin{equation}
\label{eq: vs}
{\bf E}_s = {\bf Q }_s{\bf T}
=
\begin{bmatrix}
{\bf AT} \\
{\bf A\Phi T} \\
\end{bmatrix} 
=
\begin{bmatrix}
{\bf E}_{1}\\
{\bf E}_{2} \\
\end{bmatrix} 
\end{equation}

It can be seen that the subspace of ${\bf E}_{1}$, ${\bf E}_{2}$ and ${\bf A}$ are the same. So ${\bf E}_{1}$, ${\bf E}_{2}$ and ${\bf A}$ have the same range \cite{Roy1989}. As a result, a nonsingular $\mathcal{L} \times \mathcal{L}$ matrix ${\bf \Psi}$ can be defined as
\begin{equation} 
{\bf E}_{1}{\bf \Psi} = {\bf E}_{2},
\end{equation}
hence ${\bf \Psi}$ can be defined by:
\begin{equation}
\begin{array}{rcl}
\vspace{2mm}
 {\bf AT\Psi}& = & {\bf A\Phi T} \\
\vspace{2mm}
{\bf AT\Psi T}^{-1}&  = & {\bf A\Phi} \\
\vspace{2mm}
{\bf \Psi} & = & {\bf T}^{-1}{\bf \Phi T} .
\end{array}
\end{equation}

As a result, the eigenvalues of ${\bf \Psi}$ must be equal to the diagonal elements of the ${\bf \Phi}$, and ${\bf T}$ columns are the eigenvectors of ${\bf \Psi}$. This is the key relationship in the development of ESPRIT and their properties. The signal parameters are obtained as nonlinear functions of the eigenvalues of the operator that maps ${\bf \Psi}$  one set of vectors (${\bf E}_1$) spanning an $m$-dimensional signal subspace into another (${\bf E}_2$) \cite{Roy1989,Haardt1997}. Then, since the $\mathcal{L}$ eigenvalues $\phi_\ell$ of ${\bf \Phi}$ are calculated, the angles of arrival can be computed as:
\begin{subequations}
\begin{align}
\phi_\ell &= e^{j\frac{2\pi}{\lambda} \Delta \theta_\ell}, \quad \qquad \ell=1\ldots \mathcal{L}\\
\label{eq:doa_esprit}
\theta_\ell &= {\rm arcsin}\rbrac{\frac{\lambda \cdot {\rm arg}(\phi_\ell)}{2\pi \Delta}}
\end{align}
\end{subequations}
where ${\rm arg}(\phi)=\arctan\rbrac{\frac{{\rm Im}(\phi)}{{\rm Re}(\phi)}}$. 

The ESPRIT-DoA procedure estimates a basis for the signal subspace, ${\bf E}_1$ and ${\bf E}_2$, then find ${\bf \Psi}$, next compute the eigenvalues of ${\bf \Phi}$, {\it i.e.}, $\phi_1, \phi_2\ldots \phi_\mathcal{L}$ and finally compute the DoA applying \eqref{eq:doa_esprit}. A pseudo-code for the ESPRIT-DoA procedure is described in the Algorithm \ref{ESPRIT code}, where $\mathcal{L}$ is the number of sources, $M$ the number of antennas and $S$ is the number of samples.
\begin{algorithm}
\caption{ESPRIT-DoA Procedure}
\label{ESPRIT code}
\begin{algorithmic}[1]
\State {$\vect{\hat R} =\frac{1}{S}\sum_{n=0}^{S-1}{\bf x}_n {\bf x}_n^H$} \Comment{Autocorr. Matrix - {$SM^2+M^2$}}
\vspace{1mm}
\State $\vect{\hat R} =\vect{QDQ}^H$ \Comment{Eigendecomposition - {$\frac{2M^3}{3}$}}
\vspace{1mm}
\State $\vect{\Psi}=\vect{T\Phi T}^H$ \Comment{Eigendecomposition - {$\frac{2M^3}{3}$}}
\vspace{1mm}
\State {${\bf E}_{1}{\bf \Psi} = {\bf E}_{2}$} \Comment{Matrix multiplication - {$\frac{\mathcal{L}^{2}}{2}+\frac{\mathcal{L}}{2}$}}
\vspace{1mm}
\State {Solve $\phi$} \Comment{Eigenvalues - {$\frac{2M^3}{3}$}}
\vspace{1mm}
\State {$\theta_k = {\rm arcsin}\rbrac{\frac{\lambda \cdot {\rm arg}(\phi_k)}{2\pi \Delta}}$} \Comment{Find DoA - {$\frac{\mathcal{L}^{2}}{2}+\frac{\mathcal{L}}{2}$}}
\end{algorithmic}
\Comment{Total complexity: {$M^{2}(S+\frac{6M}{3}+1)+\mathcal{L}(\mathcal{L}+1)$}}\\
\end{algorithm}

\subsubsection{Unitary-ESPRIT}
The Unitary-ESPRIT is a method derived from the classic ESPRIT \cite{haardt2014}. The main feature of this method is the real decomposition of the matrices, which reduces computational complexity. For the real transformation, let's define ${\bf \Pi}_p$ as a $p \times p$ exchanging matrix with ones on its antidiagonal and zeros elsewhere:
\begin{equation}
{\bf \Pi}_p = 
\left[ {\begin{array}{*{20}{cccc}}
 & & &1 \\
 & & 1& \\
 & \cdot & & \\
 1 & & & 
\end{array}} \right]\quad 
\in \mathbb{R}^{p \times p}
\end{equation}
Moreover, a complex matrix ${\bf M} \in \mathbb{C}^{p \times q}$ is called centro-Hermitian if
\begin{equation}
{\bf \Pi}_p {\bf M}^* {\bf \Pi}_q = {\bf M}
 \end{equation}
 where $(\cdot)^*$ is the complex conjugation without transposition. \\
 A matrix ${\bf Q} \in \mathbb{C}^{p \times p}$ is left ${\bf \Pi}$-real if satisfy:
 \begin{equation}
 {\bf \Pi}_p{\bf Q}^* = {\bf Q}
 \end{equation}\\
 The special set of unitary sparse left ${\bf \Pi}$-real matrices is denoted as ${\bf Q}_p$. They are given by
 \begin{equation}
 {\bf Q}_{2n} = \frac{1}{\sqrt{2}}
 \left[ {\begin{array}{*{20}{cc}}
{\bf I}_n & j{\bf I}_n  \\
{\bf \Pi}_n & -j{\bf \Pi}_n    
\end{array}} \right]
 \end{equation}
 and
  \begin{equation}
  {\bf Q}_{2n+1} = \frac{1}{\sqrt{2}}
 \left[ {\begin{array}{*{20}{ccc}}
{\bf I}_n & {\bf 0}_{n \times 1}  & j{\bf I}_n \\
{\bf 0} ^T_{n \times 1}  & \sqrt{2} & {\bf 0} ^T_{n \times 1} \\    
{\bf \Pi}_n & {\bf 0}_{n \times 1}  & -j{\bf \Pi}_n
\end{array}} \right]
 \end{equation}
for even and odd order, respectively.

The shift-invariance property related to the U-ESPRIT method is defined by the selection matrix:
\begin{equation}
\vect{J_2} = \sbrac{ \vect{0}_{m \times1} \quad \vect{I}_{m} } \in \mathbb{R}^{m \times M}
\end{equation}
where $m$ is the number of sensors of the subarrays. Hence,  defining real-value transformations for the selection matrix as:
\begin{equation}
\vect{K_1} = 2\, {\rm Re}\cbrac{\vect{Q}_{m}^H \vect{J_2} \vect{Q}_M}
\end{equation}
and 
\begin{equation}
\vect{K_2} = 2 \, {\rm Im}\cbrac{\vect{Q}_{m}^H \vect{J_2} \vect{Q}_M}
\end{equation}

In case the array is center-symmetric,  the forward-backward averaging (FBA) procedure can be applied to the data matrix ${\bf X}$, written as $S$ samples-matrix of \eqref{eq:array_data_vector}: 
\begin{equation}
{\bf X}=\sbrac{ {\bf x}_{n1} \, {\bf x}_{n2} \, ...\, {\bf x}_{nS}}.
\end{equation}
 Indeed, the FBA procedure uses the symmetry of the data to create an additional set of $S$ virtual samples. Also, via FBA, two coherent sources can be decorrelated. The signal matrix $\vect{X}_{\rm fba}$ is defined as:
\begin{equation}
\vect{X}_{\rm fba} = \sbrac{\vect{X} \quad \vect{\Pi}_M\vect{X}^*\vect{\Pi}_S} \in \mathbb{C}^{M \times 2S}
\end{equation}

\begin{figure}[!htbp]
\centering
\includegraphics[width=0.75\textwidth]{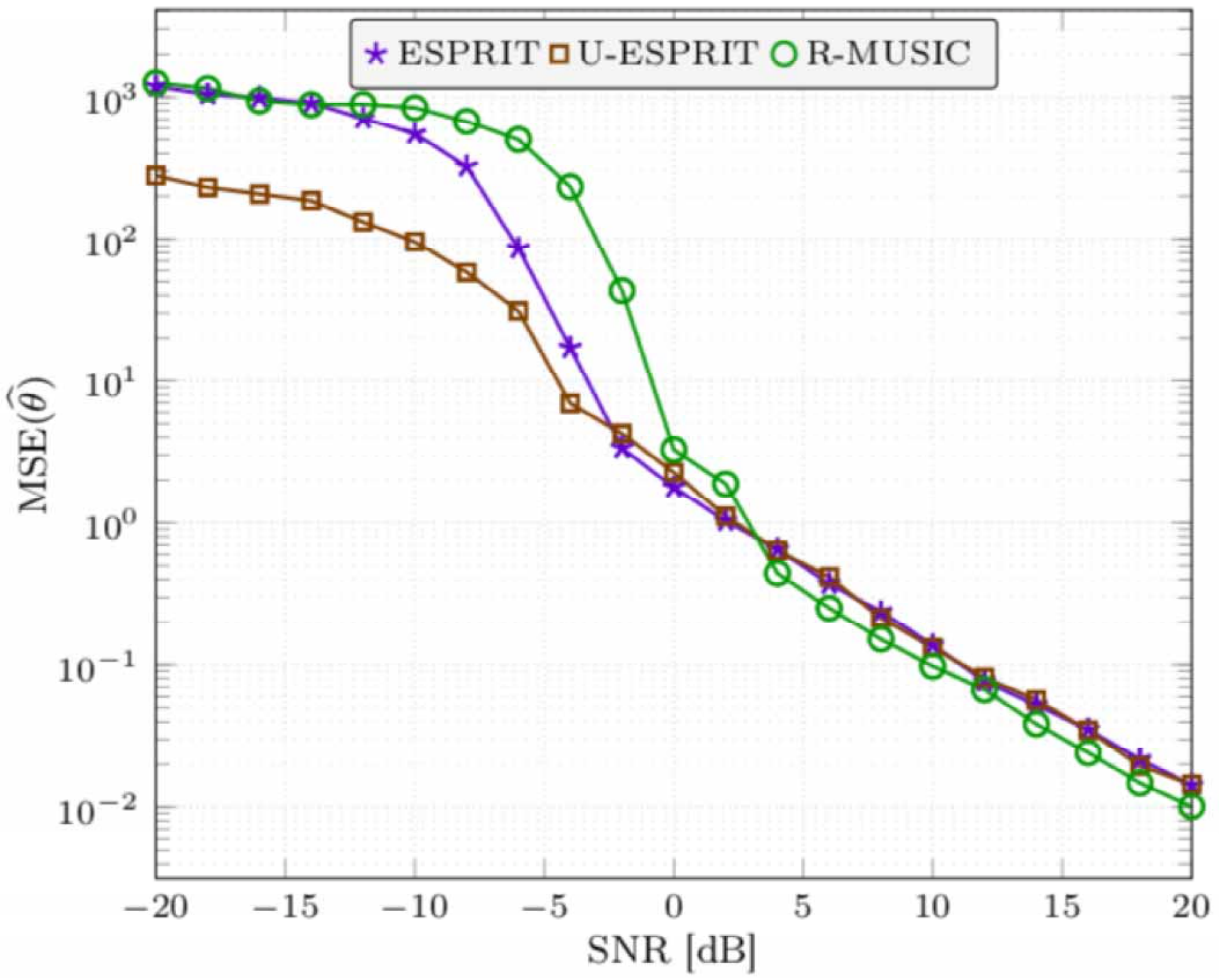}
\caption{MSE {\it vs} SNR for the DoA with $\mathcal{L}=2$, $M=8$ antennas, $S=16$ samples, $\theta \in \{10^\circ;\, 15^\circ \}$.}
\label{fig:mse_haardt2014}
\end{figure}

Briefly, the U-ESPRIT method is composed of 3 steps:
\begin{itemize}
\item Estimation of the real subspace.
\item Solution of least squares problem.
\item Final decomposition of eigenvalues (EVD).
\end{itemize}
The procedure to compute the U-ESPRIT-DoA is described in Algorithm \ref{UESPRIT code}. Besides, Fig. \ref{fig:mse_haardt2014} compares the performance attained by the conventional and unitary ESPRIT {\it vs} MUSIC methods obtained in a simple DoA scenario aiming at highlighting the better mean square error (MSE) performance of  the U-ESPRIT  \cite{haardt2014} in low signal-to-noise ratio (SNR) regime. 
\begin{algorithm}
\caption{U-ESPRIT-DoA Procedure}
\label{UESPRIT code}
\begin{algorithmic}[1]
\Require $\vect{X}$ and $\mathcal{L}$.
\State Compute {$\vect{X}_{\rm fba}$}  \Comment{$2M^3+MS^2$} %
\vspace{1mm}
\State {$\vect{\widehat R} =\frac{1}{S}\sum_{n=0}^{S-1}\vect{X}_{\rm fba}. \vect{X}_{\rm fba}^H$} \Comment{$SM^2+M^2$}%
\vspace{1mm}
\State $\mathcal{T} = {\rm Re}\cbrac{\vect{Q}_M^H \vect{\widehat R} \vect{Q}_{M}}$  \Comment{{$2M^2$}}%
\vspace{1mm}
\State Compute $\vect{\widehat{E}_s}$ from SVD of $\mathcal{T}$ \Comment{$\frac{2M^3}{3}$} %
\vspace{1mm}
\State Compute $\vect{{K_1}}$ and $\vect{{K_2}}$ \Comment{$4M^2$}%
\vspace{1mm}
\State {$ {\bf {K_1}}\vect{\widehat{E}_s}\mathbold{\Gamma} \approx {{\bf K_2}}\vect{\widehat{E}_s} $} \Comment{$2M^2$} %
\vspace{1mm}
\State {$\mathbold{\widehat{\Gamma}}_{\rm LS} = (\vect{{K_1}}\vect{\widehat{E}_s})^\dagger (\vect{{K_2}}\vect{\widehat{E}_s})$}  \Comment{$M^3+2M^2+M$}%
\vspace{1mm}
\State {Solve the eigenvalues $\phi$ of $\mathbold{\widehat{\Gamma}}_{\rm LS}$}  \Comment{{$\frac{2M^3}{3}$}} %
\vspace{1mm}
\State {$\theta_k = {\rm arcsin}\rbrac{\frac{2.{\rm arctan}(\phi_k)}{\frac{2\pi \Delta}{\lambda}}}$} \Comment{{$\frac{\mathcal{L}^{2}}{2}+\frac{\mathcal{L}}{2}$}}
\end{algorithmic}
\Comment{Total complexity: {$\frac{13M^3}{3}+11M^2+M\rbrac{S^2+S+M+1}+\frac{\mathcal{L}^2}{2}+\frac{\mathcal{L}}{2}$}}\\
\end{algorithm}

%
\subsection{Polynomial-Rooting Techniques}
Many improvements and modifications introduced in the MUSIC algorithm have been proposed aiming to increase the resolution while trying to reduce complexity.  One of these improvements is the root-MUSIC algorithm. Polynomial-rooting techniques use a polynomial parameterization to estimate DoA, where the roots of a polynomial are the estimated angles. There are several techniques that use this approach, however for comparison purpose with the MUSIC and other methods, herein we have chosen  Root-MUSIC method. Such technique has been selected for analysis and comparison since it is directly derived from the conventional MUSIC algorithm while achieves good performance-complexity tradeoffs; hence, despite similarities, the Root-MUSIC results in low-complexity and simple implementation appeal.

The root-MUSIC algorithm was proposed by Barabell \cite{Barabell1983}; it is based on the polynomial rooting approach and provides improved resolution regarding the classical MUSIC method, especially at low SNR regime. As previously defined in Eq. (\ref{eq:musicequation}), the MUSIC spatial spectrum is expressed by \cite{Barabell1983,BenAllen2006}:
\begin{equation}
\label{eq:musicequation_root}
{\bf P_\textsc{music}}(\theta)  =  \frac{1}{{\bf a}^{H}(\theta)\bar{\bf A}{\bf a}(\theta)}  
\end{equation}
where $\bar{\bf A}= {\bf Q}_{n}{\bf Q}_{n}^{H}$. 
Rewriting \eqref{eq:musicequation_root}, we can obtain
\begin{equation}
{\bf P}_{\textsc{r-music}}^{-1} = \sum_{m=1}^{M} \sum_{n=1}^{M} e^{-j\frac{2\pi}{\lambda}md\sin\theta}\bar{\rm A}_{mn} e^{j\frac{2\pi}{\lambda}nd\sin\theta}
\end{equation}
or simplified to 
\begin{equation}
{\bf P}_{\textsc{r-music}}^{-1} = \sum_{l = -M +1}^{M-1}\bar{a}_{l} e^{-j\frac{2\pi}{\lambda}ld\sin\theta}
\end{equation}
where $\bar{a}$ is the sum of entries of $\bar{\rm A}$ along the $l$th diagonal, i.e,
\begin{equation}
\bar{a}_l \overset{\Delta}{=}  \sum_{m-n=l} \bar{\rm A}_{mn}.
\end{equation}

Now defining a $2(M+1)$ order polynomial $G(z)$ \cite{Barabell1983}:
\begin{equation}
G(z) = \sum_{l= -M +1}^{M+1}\bar{a}_{l} z^{-l}, 
\end{equation}
by resolving the $G(z)$ on a unit circle, the MUSIC spectrum can be evaluated \cite{Barabell1983,BenAllen2006}. The roots of $G(z)$ close to the unit circle are equivalent to MUSIC peaks, {\it i.e.}, the $l$th pole of $G(z)$ at $z=z_{l}=\abs{z_{l}}e^{j{\rm arg}(z_{l})}$ conducing to: 
\begin{equation}
\theta_{l} = \arcsin\rbrac{\frac{\lambda\cdot {\rm arg}(z_{l})}{2\pi d}}
\end{equation}
A pseudo-code for the Root-MUSIC-DoA is presented in Algorithm \ref{rmusic code}.
\begin{algorithm}
\caption{Root-MUSIC-DoA Procedure}\label{rmusic code}
\begin{algorithmic}[1]
\State {$\vect{\hat R} =\frac{1}{S}\sum_{n=0}^{S-1}{\bf x}_n {\bf x}_n^H$} \Comment{Autocorr. Matrix - {$SM^2+M^2$}}
\vspace{1mm}
\State $\vect{\hat R} =\vect{QDQ}^H$ \Comment{Eigendecomposition - {$\frac{2M^3}{3}$}}
\vspace{1mm}
\State {$\bar{\bf A} = {\bf Q}_{n}{\bf Q}_{n}^{H}$} \Comment{Eigenvectors multiplication - {$M^3+M^2\mathcal{L}$}}
\vspace{1mm}
\State Solve $G(z) $ \Comment{Find polynomial roots - {$2M^{2}(M-1)$}}
\vspace{1mm}
\State $\theta_l = \arcsin\rbrac{\frac{{\rm arg}(z_{l})}{\frac{2\pi d}{\lambda}}}$ \Comment{Find angles - {$2(M-1)$}}
\end{algorithmic}
\Comment{Total complexity: $\frac{11}{3}M^3+M^2(S+\mathcal{L}-1)+2(M-1)$}
\end{algorithm}

\subsection{FT-DoA Method}
In this section, we discuss the method based on the spatial spectrum analysis of the signals received to estimate the angle of arrival. The energy distribution in the space domain is obtained by spectral analysis (Fourier Transform, FT) and is used for DoA estimation. 

Based on the steering matrix ${\bf A}$ described in (\ref{eq:a_matrix}), we define
\begin{equation}
u = \frac{d}{\lambda} \cos \theta
\end{equation}
One can associate the antenna array with a sampling system to take samples of signals received in the spatial domain. The space between elements of the array $d$ is associated with the sampling period; hence, $u$ is associated with the spatial frequency at the elevation angle $\theta$ \cite{Zhang2011b}.

Notice that the FT is deployed to obtain the frequency domain signal from the time signals; hence, 1D-FT on $u$ is applied to obtain the spatial spectrum of the received signals. Regarding $u$, the spatial spectrum of the received signals is defined by
\begin{equation}
\label{eq:spatial_spectrum}
{\bf x}_{u} = \sum^{M-1}_{m=0} {\bf x}_{m} e^{-j 2\pi k u}, \qquad {\bf x}_{m} = \int_{-d/\lambda}^{d/\lambda} {\bf x}_{u} e^{j 2\pi k u} du
\end{equation}

The spatial spectrum defined in (\ref{eq:spatial_spectrum}) describes the energy distribution of signals received in the spatial domain; through this information it is possible to estimate the DoA \cite{Zhang2011b}.

The spatial spectrum function defined in (\ref{eq:spatial_spectrum}) is continuous in $u$; herein, the discretized version of the spatial spectrum with respect to $u$ is defined by:
\begin{equation}
\label{eq_fft}
{\bf x}_{l} = \sum^{M-1}_{m=0} {\bf x}_{m} e^{-j 2\pi {ml} \Delta u}, \quad {\bf x}_{m} = \frac{1}{M} \sum^{M-1}_{m=0} {\bf x}_{l} e^{j 2\pi ml \Delta u}
\end{equation}
where $\Delta u = \frac{1}{M}$ is the sampling interval on $u$ in the principal period, $l$ is the serial number of the sample point on $u$. Then the $\ell$th angle of arrival can be defined as
\begin{equation}
\theta_\ell = {\rm arcsin} \rbrac{\frac{\lambda \cdot u_\ell}{d}}
\end{equation}

Deploying Fast Fourier Transform (FFT) to compute the  spatial spectrum of the received signals, one can padding zeros up to length of FFT, $N_\textsc{fft}$, to improve performance of the FT-DoA method. Hence, in (\ref{eq_fft}) only one snapshot of the signal is need, while we can let even compute $S$ times and calculate the mean to improve the estimation.  A pseudo-code for FT-DoA is depicted in Algorithm \ref{FT code}.

\begin{algorithm}
\caption{FT-DoA Procedure}
\label{FT code}
\begin{algorithmic}[1]
\Require Matrix of received signals $\vect{X}$.
\State Compute the FFT of {$\vect{X}$} \Comment{$SN_{FFT}{\rm log}(N_{FFT})$} %
\vspace{1mm}
\State Findpeaks \Comment{$4\mathcal{L}N_{FFT}$}
\vspace{1mm}
\end{algorithmic}
\Comment{Total complexity: {$\rbrac{N_{FFT}{\rm log}(N_{FFT})+4\mathcal{L}N_{FFT}}S$}}\\
\end{algorithm}

\section{Performance-Complexity Analysis}\label{sec:DoA_compare}
In this section the performance {\it versus} complexity tradeoff of various DoA methods discussed previously is compared by means of figure-of-merits, namely {\it DoA error} and  {\it scattering of estimation}. In this sense, the error of amplitude of the scanning methods DS, MVDR, MUSIC and FT-DoA, as well as the DoA error of all methods are analysed. Table \ref{tab:setup} summarizes the main parameters values adopted in this section.
\begin{table}[H]
\caption{Adopted Parameters Values and Metrics}
\vspace{-2mm}
\begin{center}
\begin{tabular}{lll}
\hline
\bf Parameter &&\bf Values\\
\hline
Antenna array type   && linear; equidistant elements\\
Number of sources && $\mathcal{L}\in\{1; 13\}$\\
Number of element antennas&& $M\in\{2;\ldots 32;\ldots; 256\}$\\
Number of samples && $S=1000 $\\
{Number of FT samples} && {$N_{\rm FFT}=1024$}\\
Signal-to-noise ratio & & SNR$\in\{-20;\, 0;\, 5\}$ dB\\
True DoA angles && $\theta \in \{-60^\circ:10^\circ:60^\circ\; ; 85^\circ\}$\\
\hline
\multicolumn{3}{c}{\bf Estimation Process}\\
\hline
Number of  Realizations && $\mathcal{I} \in \{1;10 ;200\}$\\
\hline
\multicolumn{3}{c}{\bf DoA Method Parameters}\\
\hline
\textsc{esprit} subarray formation && $m=M-1$\\
\hline
\end{tabular}
\end{center}
\label{tab:setup}
\end{table}

\subsection{Magnitude Amplitude Analysis}

\begin{figure}[!htbp]
\centering
\includegraphics[width=.99\textwidth]{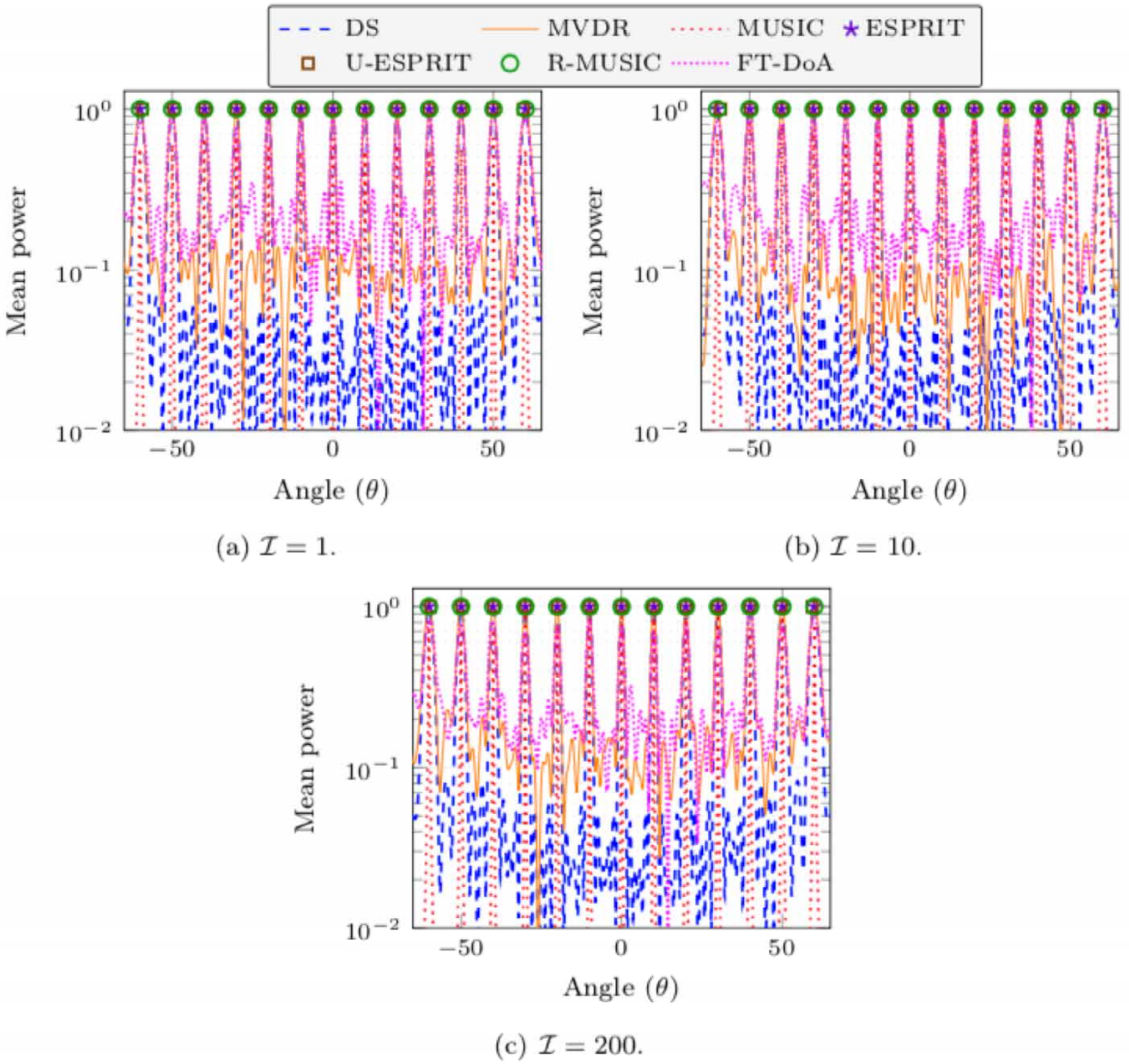}
\caption{Mean power for the DoA methods under $\mathcal{L}=13$ equidistant sources, $M=64$ antennas, and low SNR$=-20$dB.} 
\label{fig:mean_i}
\end{figure}

Considering that the output power of the DS, MVDR, MUSIC and FT-DoA method direction of arrival estimators is the mean power,  it is reasonable to take the average of several realizations. Let's consider $\mathcal{I}$ realizations, $\mathcal{L}$ sources with directions in the range $\theta \in \{-60^\circ:10^\circ:60^\circ\}$, $M$ antennas{, $N_{\rm FFT}$ FT samples} and $S$ samples, we performed numerical simulations within $\mathcal{I} \in \{1;\, 10;\, \ldots;\, 200\}$ realizations. The results are depicted in Fig. \ref{fig:mean_i}, considering average behavior over a low medium and high number of realizations, i.e., $\mathcal{I}=1, 10$ and $200$, respectively. Notice that with the increase of realizations, the power differences in the DoA localization outputs become more and more smooth, mainly in the MUSIC method. 
%
%
This behavior is due to the stochastic characteristics of the methods. Notice that the ESPRIT{, U-ESPRIT} and Root-MUSIC methods do not present amplitude variation since they do not deploy exhaustive search procedure. This analysis was performed to demonstrate the behavior of the estimators taking into account the average estimation over several realizations. {Even with one realization, it is possible to estimate DoA, however for applications that also require the power information of the received signal, it is essential to take into account the average of the realizations.}

\subsection{DoA Error}
The mean square error (MSE) of the estimated DoA $\widehat{\theta}$ for the DS, MUSIC, MVDR, ESPRIT, {U-ESPRIT}, Root-MUSIC {and FT-DoA} methods is strainghtforward computed as: 
\begin{equation}
\begin{array}{ccl}
\vspace{1mm}
\textsc{mse}(\widehat{\theta}) & = & \expect{\varepsilon} \,  \,= \,\,\expect{\sum_{{i}=1}^{\mathcal{I}}\abs{\theta_{{i}} - \theta_{\rm true}}^2}\\
\vspace{1mm}
& \approx & \frac{1}{\mathcal{I}-1}\sum_{{i}=1}^{\mathcal{I}} \varepsilon_{{i}}
\end{array}
\end{equation}
where $\theta_{{i}}$ is the current estimated angle at ${i}$th realization, $\theta_{\rm true}$ is the true angle-of-arrival of the signal source and $\mathcal{I}$ is the number of realizations of the DoA estimation method.

\begin{figure}[!htbp]
\centering
\includegraphics[width=0.98\textwidth]{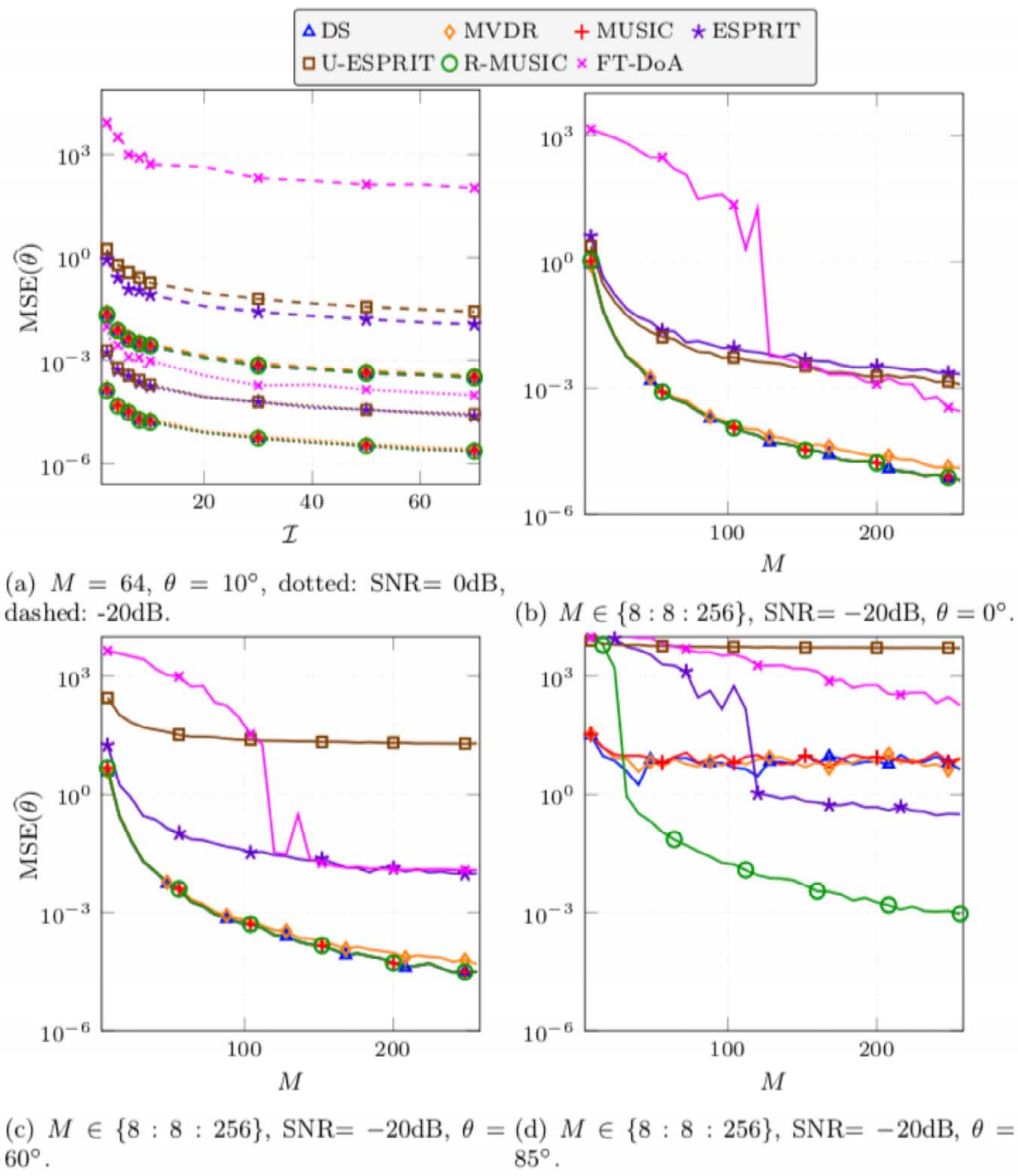}
\caption{\textsc{mse} for the DoA methods; $\mathcal{L}=1$, $S=1000$ samples,$N_{\textsc{fft}}=1024$.}
\label{fig:mse}
\end{figure}

Fig. \ref{fig:mse}(a) illustrates the \textsc{mse} for a wide range of number of realizations, $\mathcal{I} \in[2;\, 70]$ and for both low and medium-high SNR, {\it i.e.,} SNR $\in \{-20; \, 0\}$ dB. One can see that adopting $\mathcal{I}=30$ iterations is enough for all analized DoA methods to achieve {very close to} their asymptotic \textsc{mse} performance condition.

Besides, Fig. \ref{fig:mse}(b), Fig. \ref{fig:mse}(c) and Fig. \ref{fig:mse}(d), illustrate the \textsc{mse} for a wide range of number of antennas, $M \in\{8:8:256\}$ at various angles to cover the entire possible range of DoA estimation, {\it i.e.}, $0<|\theta| \in \{ 0^{\circ}; 60^{\circ}; 85^{\circ} \} < 90^{\circ}$ and $\mathcal{I}=40$ realizations, since the numerical results in Fig. \ref{fig:mse}(a)  indicated that a number of realizations $\mathcal{I}\geq30$ is enough to attain the asymptotic \textsc{mse}.  All the methods have presented improvements with the increase of the number of element antennas $M$ and in the range of angles up to $|\theta|< 60^{\circ}$; besides, all of them showed a well behaved operation. {The FT-DoA method presented a considerable reduction of the MSE to $M > 100$; this is because in this method the samples are obtained by the antennas so for reduced numbers of antennas this method performs worse than the others. It is observed that this method does not reach the low MSE levels like the others. Though, the levels are small enough for most applications.} However, in the range near to the limit, {\it i.e.} $85^{\circ}$, the search methods demonstrated an accuracy reduction in the  estimations, despite the ESPRIT and Root-MUSIC have reasonable operation for high number of antennas, while the Root-MUSIC achieves the best performance in such limit scenario. For instance, it can be seen that with $M \geq 100$ the \textsc{MSE} is much smaller than for $M \leq 10$. This demonstrates the potential of improvement in terms of \textsc{MSE} performance from these methods in applications with \mm antennas scenarios.

\begin{figure}
    \centering
    \begin{subfigure}[b]{0.485\textwidth}
        \includegraphics[width=\textwidth]{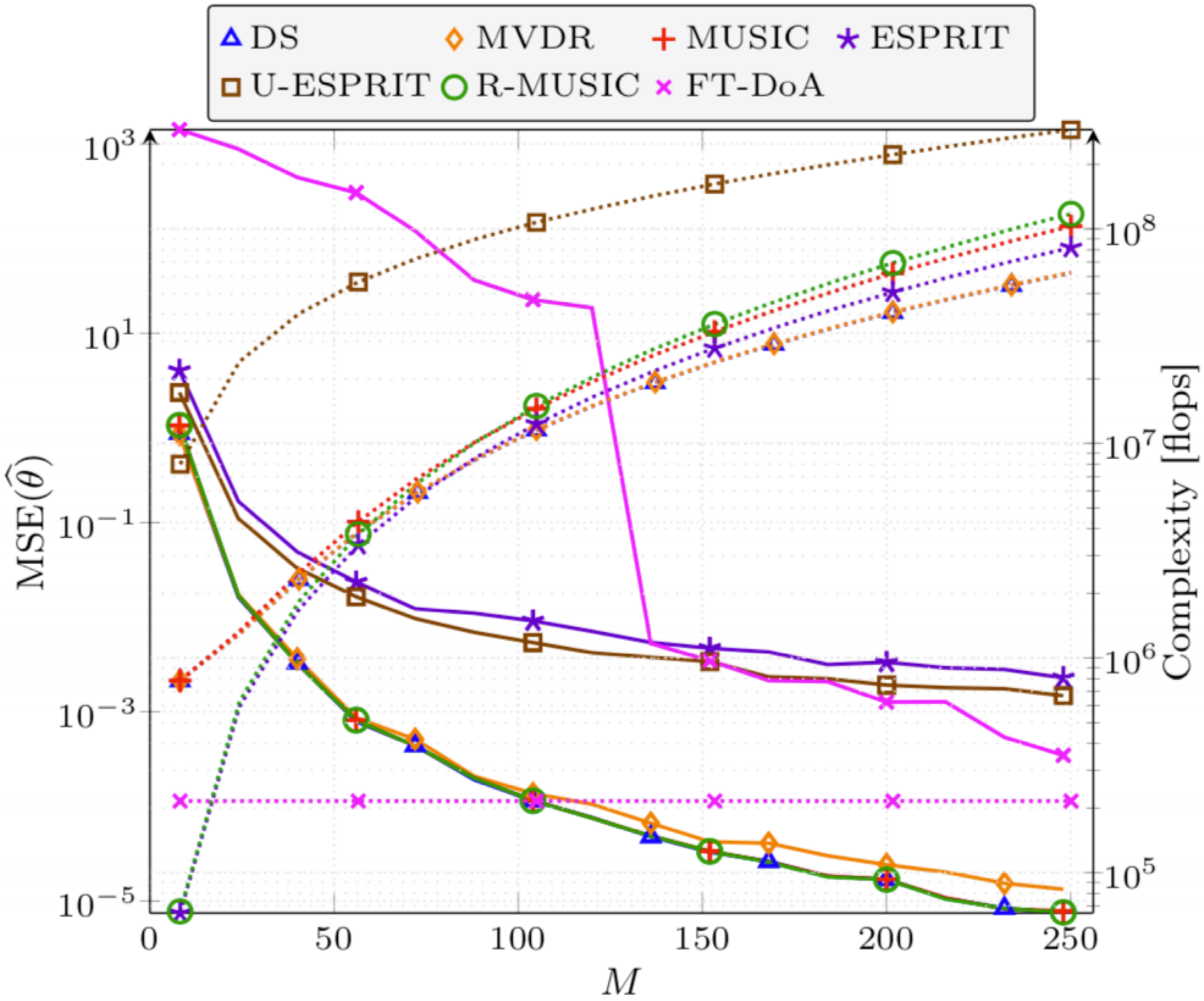}
	\caption{MSE $\times$ Complexity considering $\mathcal{L}=1$, {$M\in \{{8:8:256}\}$}, SNR$=-20$dB, $S=1000$, {$N_{\textsc{fft}}=1024$}; solid lines: MSE, dashed lines: flops complexity. \vspace{2mm}}
	\label{fig:mse_cplx}
    \end{subfigure}
    \begin{subfigure}[b]{0.485\textwidth}
        \includegraphics[width=\textwidth]{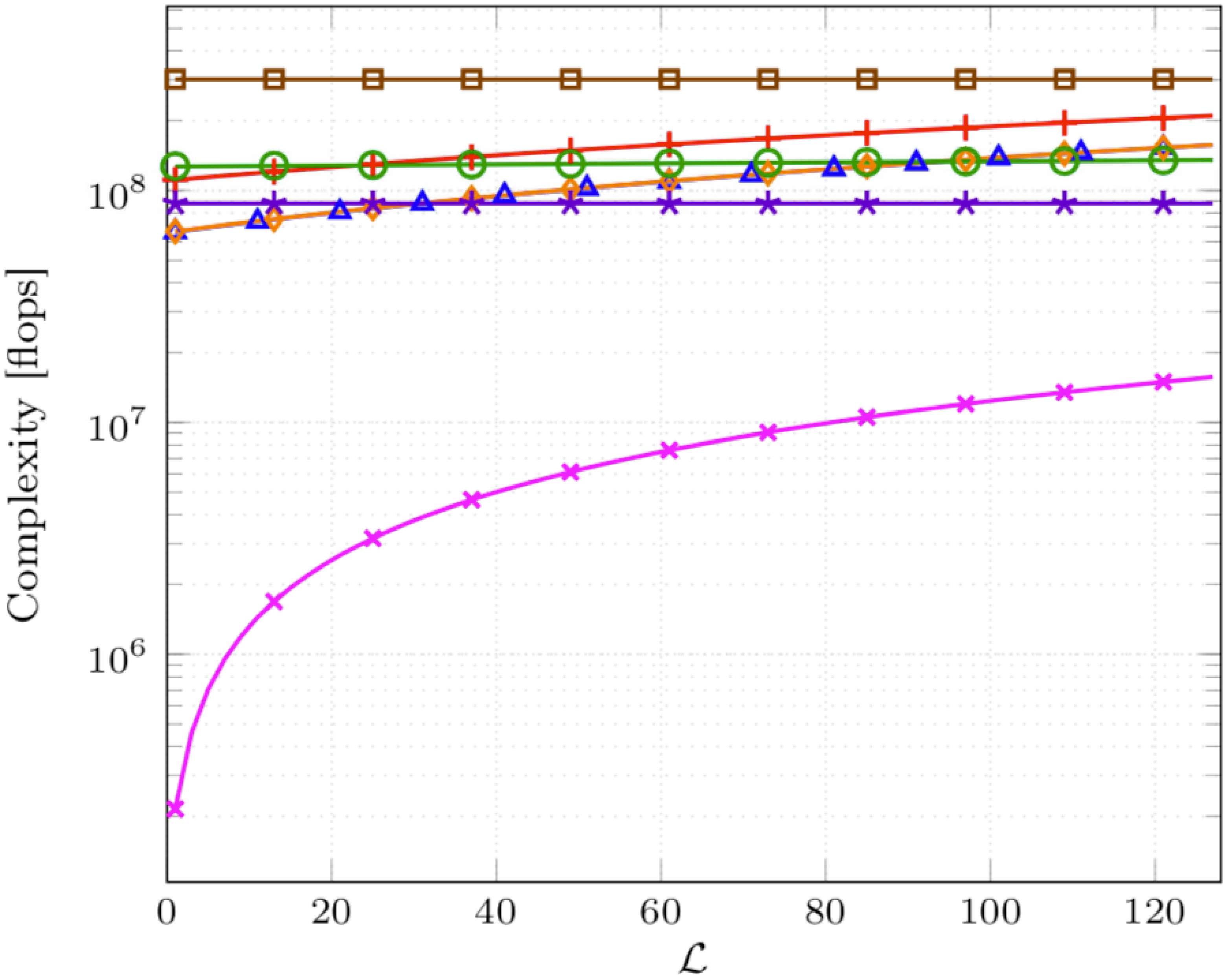}
	\caption{Computational complexity; {$\mathcal{L} \in \{ 1:1:128\}$}, ${M=256}$, $S=1000$ samples, number of scan steps of $\theta$, {\it i.e.} $P \in \{-90:0.001:90\}$, {and $N_{\textsc{fft}}=1024$}.}
        \label{fig:complexity.k}
    \end{subfigure} 
    
    \begin{subfigure}[b]{0.485\textwidth}
        \includegraphics[width=\textwidth]{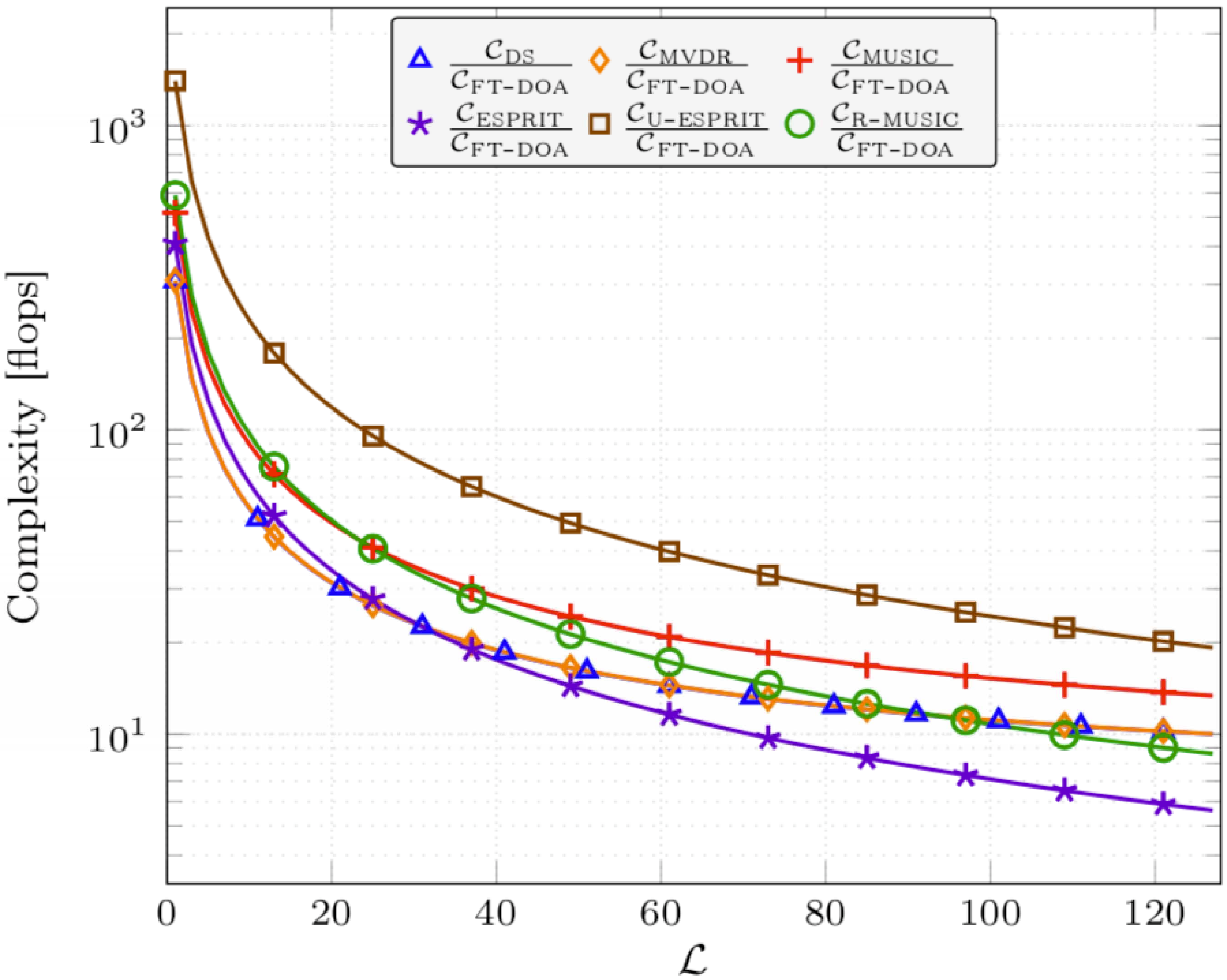}
	\caption{Relative computational complexity; {$\mathcal{L} \in \{ 1:1:128\}$}, ${M=256}$, $S=1000$ samples, number of scan steps of $\theta$, {\it i.e.} $P \in \{-90:0.001:90\}$, {and $N_{\textsc{fft}}=1024$}.}
        \label{fig:complexity.relative}
    \end{subfigure}
    \caption{MSE $\times$ Complexity and Complexity $\times$ $\mathcal{L}$ }
\end{figure}

Fig. \ref{fig:mse_cplx}  depicts the \textsc{mse} $\times$ float point operations (flops) complexity for a wide range of number of antennas, $M \in\{8:8:256\}$, $\mathcal{L}=1$ signal source, SNR$=-20$dB and $\mathcal{I}=40$ realizations. As expected, the \textsc{MSE} decreases with the number of antennas. The complexity of the ESPRIT and R-MUSIC methods is lower for approximately $M\leq50$ (crossing complexity point); while the FT-DoA method has the lowest complexity of all, and remains at the same level with the increase of the number of antennas; such feature is very interesting for \mm applications. 

As expected, the complexity of methods attain the same trend, with a very close rate of increase, except for the FT-DoA method which presents no variation of complexity with the increasing of the number of element antennas; on the other hand, the U-ESPRIT DoA method presents the greatest complexity. However, for a scenario with several sources, the complexity of the MUSIC, MVDR, DS and U-ESPRIT DoA methods increases much more than ESPRIT and R-MUSIC, which do not expand their complexities with the increase of the number of sources $\mathcal{L}$.  Besides, the FT-DoA method presents the lowest complexity as depicted in Fig. \ref{fig:complexity.k};  with such feature, it has been adopted as reference in Fig. \ref{fig:complexity.relative} and discussed in details in subsection \ref{sec:complexity}. 
%
%
\subsection{Scattering of Estimation}
Considering the scattering or dispersion of the estimation as a relevant figure-of-merit,  numerical results for the five DoA methods are discussed in this subsection aiming at establishing a comprehensive analysis on the accuracy of the DoA methods. Hence, we define the "DoA discrimination'' figure-of-merit. Due to the difference in operation of the methods, two ways were deployed to calculate DoA discrimination, one for search-based DoA methods (DS, MVDR, MUSIC and FT-DoA) and another for ESPRIT, U-ESPRIT and R-MUSIC. In the first, we adopted the 3dB power output decaying as the analysis parameter, defined by:
\begin{equation}
\mathcal{D}_{\rm 3dB}^\textsc{doa} = \left.\theta_1 - \theta_2\right| _{\Delta P}  \qquad [^\circ]
\end{equation}
where  $\Delta P = \frac{P(\theta_1)}{P(\theta_{\max})}=\frac{P(\theta_2)}{P(\theta_{\max})}=\frac{1}{2}$ and $P(\theta_{\max})=\max[P(\theta)]$. 

For the ESPRIT, {U-ESPRIT} and R-MUSIC methods, several experiments with different number of realizations $\mathcal{I}$ were carried out and the {standard deviation} between the estimated angles and the true angle $\theta_{\rm true}$ was taken, {\it i.e.}, the standard deviation of $\theta$ was evaluated as:
\begin{equation}
\mathcal{D}_{\Delta \theta}^\textsc{doa} = \sqrt{\frac{1}{\mathcal{I} -1} \sum_{i=1}^{\mathcal{I}} {\rbrac{\theta_{i} -\mu}^2}}  \qquad [^\circ]
\end{equation}
where $\theta_{i}$ is the instantaneous angle of arrival estimation and $\mu = \frac{1}{\mathcal{I}}\sum_{i=1}^{\mathcal{I}} \theta_i$ is the sample mean.

\begin{figure}[!htbp]
\centering
\includegraphics[width=0.65\textwidth]{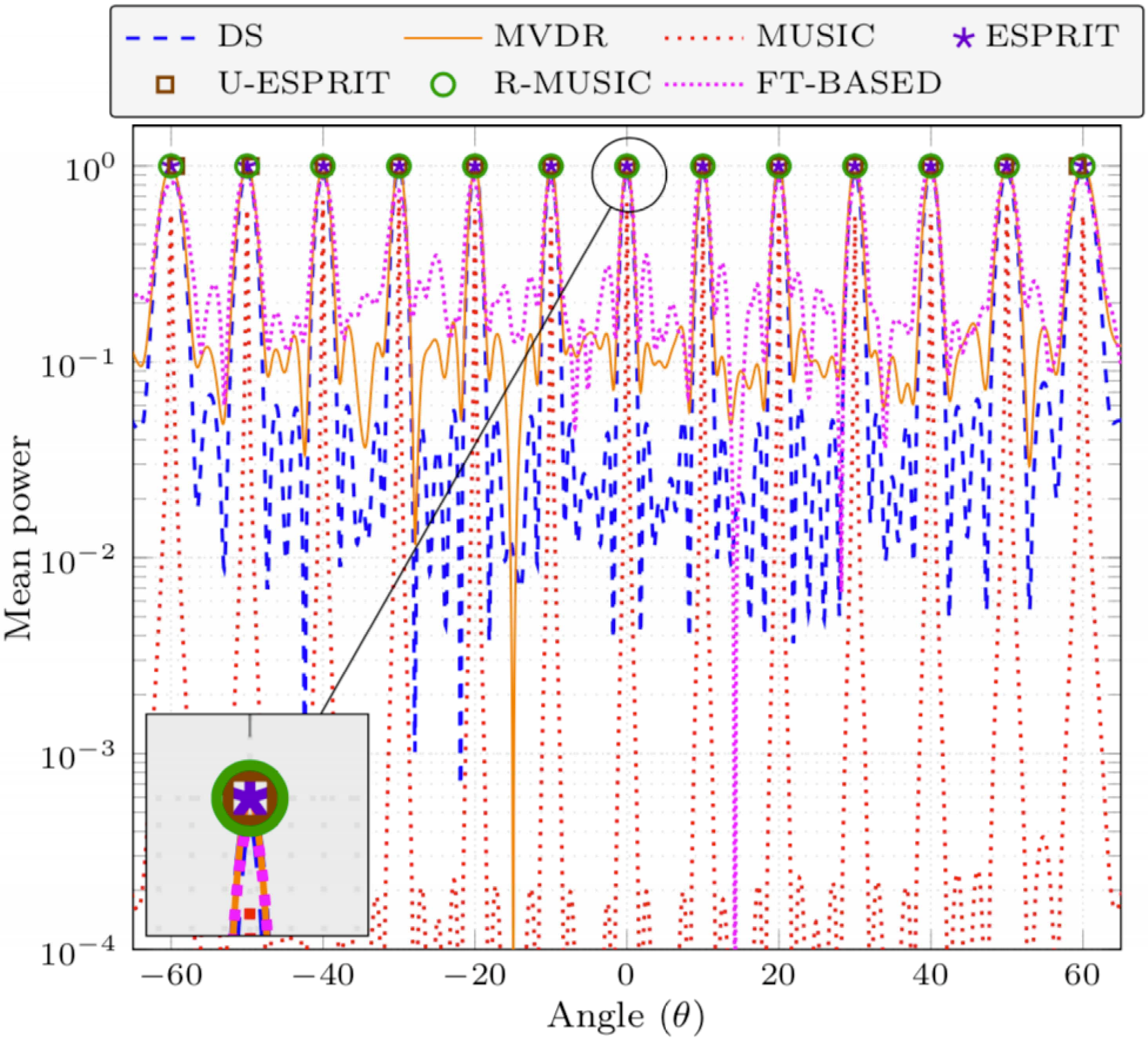}
\caption{Mean power output for the DoA methods operating under $\mathcal{L}=13$, $M=64$, SNR$=-20$ dB, $S=1000$ samples, $\theta \in \{-60^\circ: 10^\circ: 60^\circ\}$; zoom in at $\theta=0^\circ$.}
\label{fig:example_estimators}
\end{figure}

Fig. \ref{fig:example_estimators} explores the DoA estimation of multiple sources with $\theta \in \{-60^\circ:10^\circ:60^\circ\}$ for the {seven} analysed DoA estimators; in the detail is illustrated the zoom in over $\theta=0^\circ$. 
As discussed in subsection \ref{sec:complexity}, the MUSIC method and U-ESPRIT are performed with the highest computational complexity; however {MUSIC} DoA method results more accurate method, among the search methods as well, due to the loss in mean power combined to the wide scattering in DoA angle estimation.  
In contrast, there is no loss in the mean power for the other {six} methods. The MVDR presents better performance than DS, with minimized sidelobes, but with greater complexity and poor accuracy when compared with other methods. Moreover, MUSIC presents a much higher accuracy than others, with well-defined peaks, which facilitates the detection of the estimated angle, although this estimator has a higher complexity, the accuracy is high, making it very interesting method for DoA estimation. {The FT-DoA method presents similar scattering to DS and MVDR, but if the number of antennas increases, the scattering approaches to MUSIC; however, it has a very reduced complexity that will be better discussed in section \ref{sec:complexity}.} Interesting, the ESPRIT, U-ESPRIT and Root MUSIC methods do not perform the DoA estimation by the continuously calculation of the mean power across a wide range of $\theta$; instead, such methods only evaluate the estimated angle over each specific angle source, discretely. However, such method also generates a spreading in the estimation; such issue is discussed in the sequel.
%

\begin{figure}[!htbp]
\centering
\includegraphics[width=0.82\textwidth]{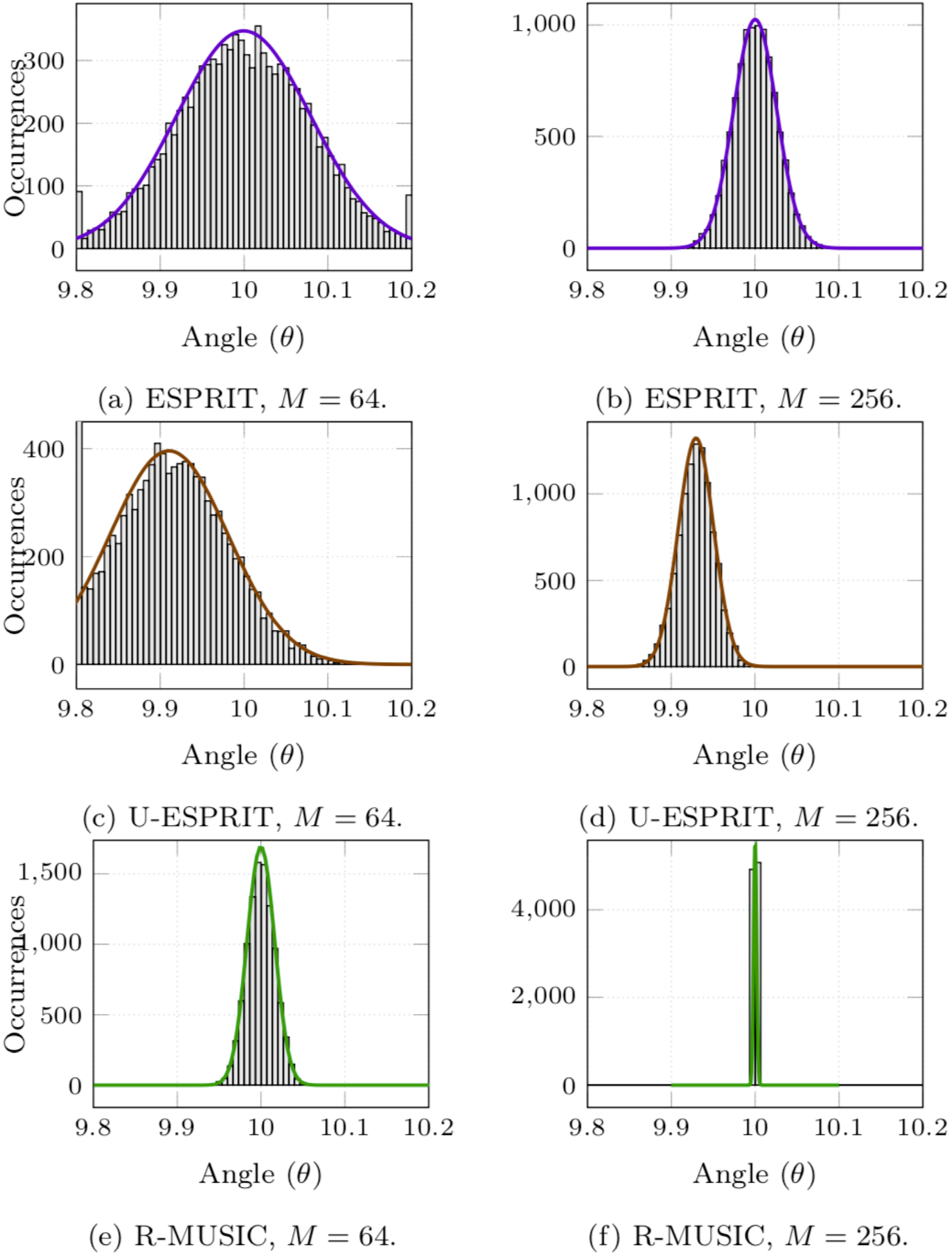}
\caption{ESPRIT, U-ESPRIT and R-MUSIC DoA distribution for $\theta_{\rm true}=10^{\circ}$, $\mathcal{I}=10^{4}$ realizations, $M \in \{ 64; 256 \}$ and SNR$=-20$dB.}
\label{fig:stat_doa}
\end{figure}

In order to better evaluate the ESPRIT, U-ESPRIT and Root-MUSIC methods, an example has been evaluated with $\mathcal{I}=10000$ realizations of a single source DoA estimation considering $\theta_{\rm true} = 10^{\circ}$, and $M \in \{64; 256\}$ element antennas. Fig. \ref{fig:stat_doa} depicts the distribution of the sample estimations while Table \ref{tab:stat_param} summarized the parameters found numerically in this example, where $\mu$ is the mean of the estimates, $\sigma$ is the standard deviation and ${\rm CI}_{95\%}$ is the 95\% confidence interval.  As one can observe, the samples follow a Gaussian distribution; indeed, all methods result in a smaller standard deviation when the number of antennas increases. The ESPRIT method shows the highest standard deviation while the R-MUSIC the lowest; the U-ESPRIT presents intermediate standard deviation values; however, regarding mean, the U-ESPRIT offers the most substantial deviation from the $\theta$-true value among the three methods.
%
%
\begin{table}[!htbp]
\caption{Synthesis of numerical analysis of ESPRIT and Root-MUSIC distribution example.}
\begin{center}
\begin{tabular}{lccccc}
\hline
{\bf DoA Method} & $\theta_{\rm true}$ & $M$ & {\bf $\mu$} & $\sigma$ &${\rm CI}_{95\%}$\\
\hline 
\textsc{esprit} & $10^{\circ}$ & 64 &  9.9940 & 0.0805 & $[9.99; 10.00]$\\
\textsc{u-esprit} & $10^{\circ}$ & 64 &  9.9104 & 0.0705 & $[9.90; 9.91]$\\
\textsc{r-music} & $10^{\circ}$& 64 & \bf 9.9998 & \bf 0.0165 & $[9.99; 10.00]$\\
\textsc{esprit} & $10^{\circ}$ & 256  & 10.0002 & 0.0257 & $[9.99; 10.00]$\\
\textsc{u-esprit} & $10^{\circ}$& 256 & 9.92966 & 0.0211 & $[9.92; 9.93]$\\
\textsc{r-music} & $10^{\circ}$ & 256  & \bf 10.0000 & \bf 0.0018 & $[9.99; 10.00]$\\
\hline
\end{tabular}
\end{center}
\label{tab:stat_param}
\end{table}

\begin{figure}[!htbp]
\centering
\includegraphics[width=0.86\textwidth]{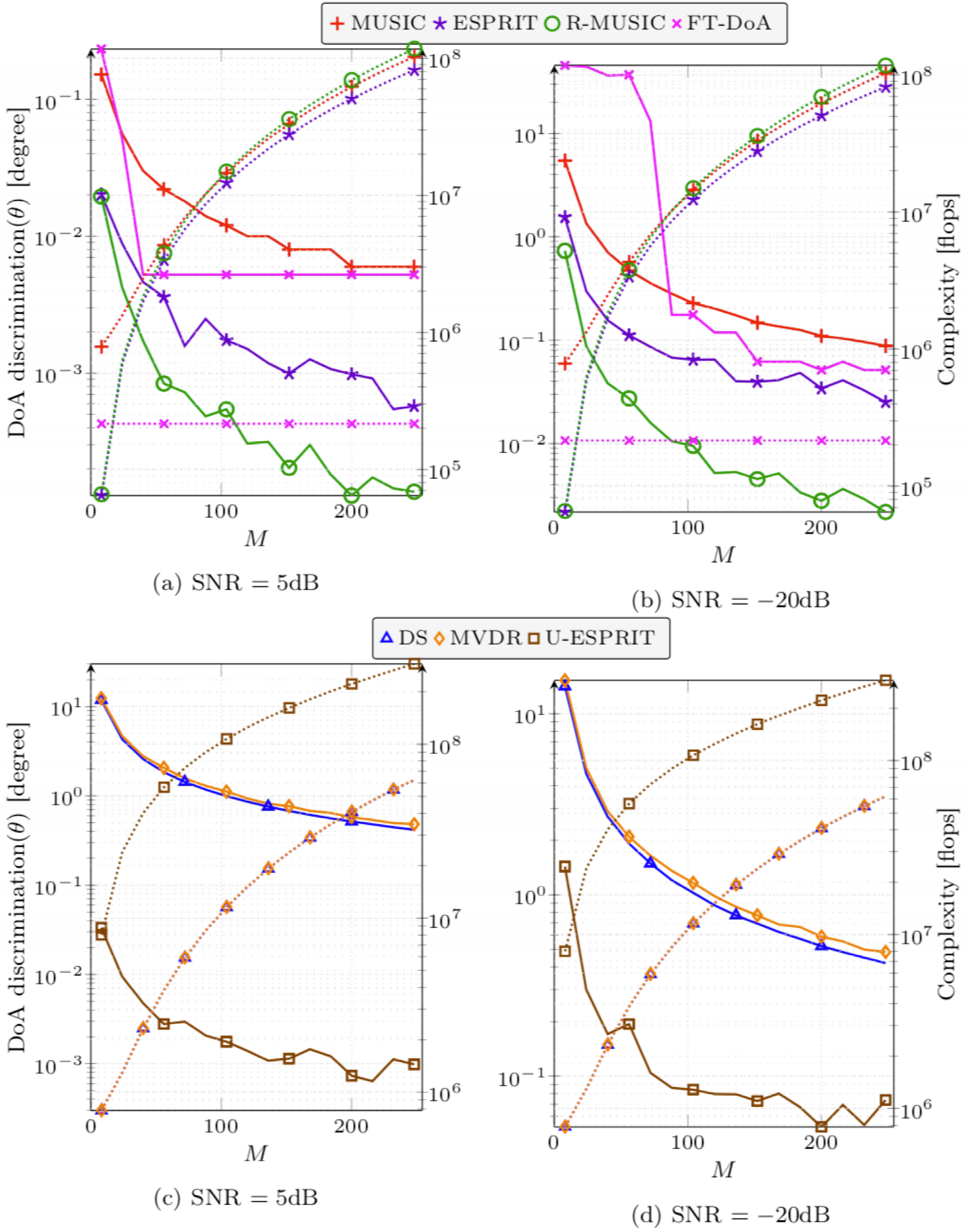}
\caption{DoA discrimination $\times$ Complexity considering $\mathcal{L}=1$, $M\in \{{8:8:256}\}$, SNR $\in \{5; \,-20\}$dB, $S=1000$, solid lines: DoA discrimination, dotted lines: Complexity.}
\label{fig:disc_cplx}
\end{figure}

Fig. \ref{fig:disc_cplx} demonstrates indubitably the superior DoA discrimination feature of the R-MUSIC method in terms of $\{ \mathcal{D}_{\rm 3dB}^\textsc{doa}; \mathcal{D}_{\Delta \theta}^\textsc{doa}\} \times M$, for low as well as high SNR regimes and wide range of number of element antennas considered. Notice that the improvement in terms of resolution or discrimination is obtained with the increase of the number of antennas, which means increased  accuracy in the DoA estimation. Besides, it is worthy to note that, under low-SNR regime, the {R-MUSIC, ESPRIT and U-ESPRIT} present the better resolution among the {seven} DoA estimators analyzed.  Furthermore, Fig. \ref{fig:disc_cplx} depicts in the left y-axis the DoA discrimination and in the righ y-axis the DoA complexity (flops) for high and low SNR regimes and different number of element antennas. Among the seven DoA methods analyzed, the R-MUSIC attains the best performance in terms of discrimination. However, in terms of complexity, the FT-DoA method presents the lowest complexity if it remains at a reasonable value for antenna numbers greater than 100{, and features DoA discrimination a bit better than MUSIC}. As a consequence, the best performance-complexity tradeoff is achieved by the FT-DoA estimator.
%
%
\subsection{Complexity}\label{sec:complexity}
Table \ref{tab:complexity} synthesizes the computational complexity in terms of number of operations{, products and summations} for the DoA methods analyzed.  The complexity of DoA methods analyzed is strongly dependent on the number of elements in the antenna array $M$, except the FT-DoA which is just dependent on the number of FFT samples ($N_{\rm fft}$). As shown in Table \ref{tab:complexity}, the MUSIC, ESPRIT, U-ESPRIT and R-MUSIC methods result in cubical polynomial complexity of order $\mathcal{O}\{M^3\}$, while DS and MVDR DoA techniques remain quadratically in complexity, of order $\mathcal{O}\{M^2S\}$; finally, the FT-DoA method has the lower complexity, of order $\mathcal{O}\{N_{\rm fft}\log N_{\rm fft}\}$.
\begin{table}[!htbp]
\caption{Computational Complexity in terms of number of operations \vspace{-6mm}}
{
\begin{center}
\scalebox{0.92}{%
\begin{tabular}{r| c| c| l}
\hline
\bf DoA & {\bf Products} & {\bf Summations} & \bf \# Operations\\
\bf Method & & & \\
\hline 
\textsc{ds}& $2M^2+4\mathcal{L}P$ & $SM^2+M$ &{$M^2(S+2)+M+4\mathcal{L}P$} \\
\textsc{mvdr}& $3M^2+M+4\mathcal{L}P$ & $2M^2+SM^2$ & {$M^2(S+6)+M+4\mathcal{L}P$}\\
\textsc{music}& $\frac{4M^3}{3}+M^2+M+4\mathcal{L}P$ & $\frac{M^3}{3}+M^2+M^2\mathcal{L}+SM^2$ &{$\frac{5}{3}M^3+M^2(S+\mathcal{L}+1)+4\mathcal{L}P$}\\
\textsc{esprit}& $M^3+M^2+\mathcal{L}^2+\frac{\mathcal{L}}{2}$ & $M^3+\frac{\mathcal{L}}{2}$ &$2M^3 + M^2\left(S+1\right)+\mathcal{L}(\mathcal{L}+1)$\\
\textsc{u-esprit}& $\frac{11M^3}{3}+10M^2+\frac{\mathcal{L}^2}{2}+MS^2$ & $\frac{2M^3}{3}+M^2+M+SM^2+\frac{\mathcal{L}}{2}$ &$\frac{13M^3}{3}+11M^2+M\rbrac{S^2+S+M+1}+\frac{\mathcal{L}^2}{2}+\frac{\mathcal{L}}{2}$\\
\textsc{r-music}& $\frac{10M^3}{3}+M^2+M^2\mathcal{L}+2M-2$ & $\frac{M^3}{3}+SM^2-2M^2$ &{$\frac{11}{3}M^3+M^2(S-1+\mathcal{L})+2(M-1)$}\\
\textsc{ft-doa}&$\rbrac{N_{\rm fft}{\rm log}(N_{\rm fft})}S$&$\rbrac{4\mathcal{L}N_{\rm fft}}S$ &{$\rbrac{N_{\rm fft}{\rm log}(N_{\rm fft})+4\mathcal{L}N_{\rm fft}}S$}\\
\hline
\end{tabular}}
\end{center}
}
\label{tab:complexity}
\end{table}

To offer a comparative view of the DoA methods complexity jointly with DoA discrimination, Fig. \ref{fig:disc_cplx} depicts the by-flops complexities (right y-axis) of all DoA methods analyzed, where $\mathcal{L}=1$ sources and number of antennas in the range $M \in \{ 8:8:256 \}$ have been considered. 
While the FT-DoA results in lowest complexity $\forall M$, the ESPRIT and R-MUSIC methods are less complex than the other remain methods, despite their close complexity when $M\geq 64$ antennas. Indeed, as the number of antennas increases, the cubic-complexity methods approaches each other, since  the number of samples $S$ remains the same or increases slowly. Indeed, in Fig. \ref{fig:disc_cplx}, as expected, up to $M\approx 64$ antennas, the R-MUSIC is quite close to ESPRIT, while MUSIC, MVDR and DS have a greater complexity, since in this range of $M$ the number of samples is preponderant. However, when the number of antennas $M \rightarrow \infty$, the complexity of the DS and MVDR results considerably smaller than the MUSIC, R-MUSIC and ESPRIT methods.

Let's analyze the relative computational complexity of the DoA methods, which has been normalized taking as reference the FT-DoA complexity ($\mathcal{C}_{\textsc{ft-doa}}$). As a result, the ratios $\frac{\mathcal{C}_{\text{DoA method}}}{\mathcal{C}_{\textsc{ft-doa}}}$, as depicted, for instance, in Fig. \ref{fig:complexity.relative}, decrease rapidly with the increasing number of signal sources in the range $\mathcal{L} \in\{ 1:1:128\}$ under a fixed number of antennas ($M=256$). Indeed, the DoA methods complexities were evaluated w.r.t. the variation of the number of sources in the range $\mathcal{L} \in \{1:1:128 \}$ under a fixed number of antennas $M=256$. As one can observe, in Fig. \ref{fig:complexity.k}, the ESPRIT, U-ESPRIT and R-MUSIC DoA methods do not present great variations in complexity with the increase of the number of sources. However, due to angle scanning feature, DS, MVDR , MUSIC {and FT-DoA techniques result in growing complexities with the increasing of the number of sources, but as expected, FT-DoA results much less complexity than the other methods, {which makes this method very promising for \mm applications.}

\subsection{{Parallelism Analysis of the DoA Algorithms}}\label{sec:parallel}
{In systems operating in real-time processing, especially the new 5G networks that require low latency, the parallel computing of the operations is crucial. In this subsection, we analyzed the parallelism of the seven DoA algorithms previously presented.}

{The principal measure of parallelization efficiency of the algorithms is the {\it Speedup}, $S_{N}$, defined by the ratio of the need time to execute the entire algorithm $\mathcal{C}$ on a single processor to the time necessary to run using $N$ processors:
\begin{equation}
S_{N} = \frac{\tau_{1}}{\tau_{N}},
\end{equation}
where $\tau_{1}$ is the time to execute the workload on a single processor and $\tau_{N}$ is the time to execute the workload on $N$ processors, being defined by:
\begin{equation}
\tau_{N} = \tau_{s} + \tau_{p} + \Delta_{\tau},
\end{equation}
where $\tau_{s}$ is the time to execute the serial portion of the workload, $\tau_{p}$ is the time to execute the parallel portion and $\Delta_{\tau}$ is an additional time due to the parallelization overhead which is quite general and accounts for any overhead due to implementing the algorithm on  a parallel way; the overhead comes from either due to the hardware, the network, the operating system, or the algorithm \cite{Amdahl1967,Culler1997,Golub2013}. It is expected that this time is a function of the number of $N$ processors deployed and the algorithm complexity $\mathcal{C}$, i.e., $\Delta_{\tau} = \Delta_{\tau}(N,\mathcal{C})$
}

\begin{figure}[!htbp]
\centering
\includegraphics[width=0.82\textwidth]{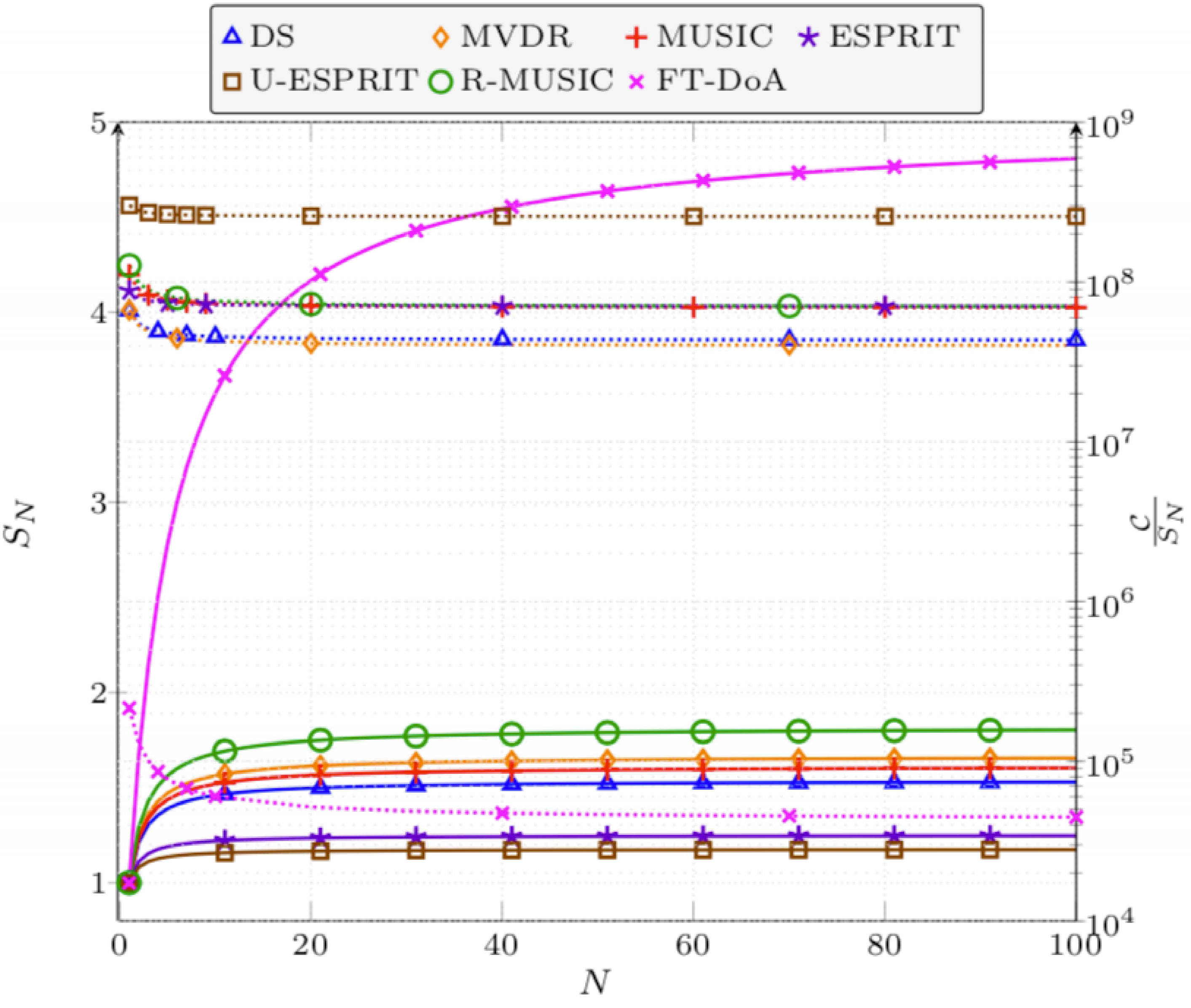}
\caption{{{\it Speedup} $\times$ ratio of Complexity and {\it Speedup} $\frac{\mathcal{C}}{S_{N}}$  considering $\mathcal{L}=1$, $M=256$, $N\in \{{1:1:100}\}$, solid lines: $S_{N}$, dotted lines: $\frac{\mathcal{C}}{S_{N}}$.}}
\label{fig:speedup}
\end{figure}

{Fig. \ref{fig:speedup} depicts the {\it Speedup} of the DoA algorithms (solid lines) as a figure-of-merit to analyze how much faster the execution of the algorithms can be with the parallel processing application. Also, the ratio between Complexity and {\it Speedup} $\rbrac{\frac{\mathcal{C}}{S_{N}}}$ of the methods is shown in the right y-axis and dotted lines). The results were computed considering $M=256$ antennas, $\mathcal{L}=1$ source and $N\in \{{1:1:100}\}$ processors. U-ESPRIT and ESPRIT have the worst $S_{N}$ performance while DS, MVDR, MUSIC, and R-MUSIC have intermediate performance; finally, the FT-DOA resulted the best performance, achieving up to 4.8 times faster processing with 100 processors.}
%
\section{Conclusions}\label{sec:concl}
Considering the seven DoA methods evaluated within the specific but representative scenarios, {\it i.e.} considering up to $\mathcal{L}=13$ sources spaced at $10^\circ$ from each other, low and medium-high SNR$ \in \{-20,\,\, 5\}$ dB and an affordable number of {$\mathcal{I}={40}$} realizations, one can conclude that the best performance-complexity tradeoff is achieved by the FT-DoA method, while the best accuracy has been achieved with Root-MUSIC DoA method. {Table \ref{tab:resumo_resul} contains a summary of the main figure-of-merit analyzed; we adopted $M=256$ antennas, SNR $\in \{-20; \,5\}$dB, $\mathcal{L} = 1$ source, ${\rm MSE}$, DoA discrimination $\mathcal{D}$ and computational complexity $\mathcal{C}$ (for  $\mathcal{L} = 1$ and $\mathcal{L} = 128$), {\it Speedup} with $N=24$ processors for all methods.}

\begin{table}[!htbp] 
\caption{{Synthesis of numerical analysis of all DoA methods analyzed.}\vspace{-5mm}}
{
\begin{center}
\scalebox{0.9}{%
\begin{tabular}{l |c| c| c| c c c c c}
\hline
{\bf DoA Method} & $M$ & $\mathcal{L}$ & SNR & ${\rm MSE}$ & $\mathcal{D}$ & $\mathcal{C} \rightarrow \mathcal{L}=1$ & $\mathcal{C} \rightarrow \mathcal{L}=128$ & $S_{N} \rightarrow N=24$  \\
\hline 
\textsc{DS} &  &  & & $6.66{\rm E}-6^{\circ}$ & $4.10{\rm E}-1^{\circ}$ & $66.30{\rm E}6$ & $157.82{\rm E}6$ & 1.50\\
\textsc{MVDR} &  &  & (-20dB) & $12.63{\rm E}-6^{\circ}$ & $4.63{\rm E}-1^{\circ}$ & $66.63{\rm E}6$ & $158.09{\rm E}6$ & 1.62\\
\textsc{MUSIC} & 256 &1 & & $6.51{\rm E}-6^{\circ}$ & $8.80{\rm E}-2^{\circ}$ & $110.47{\rm E}6$ & $194.11{\rm E}6$ & 1.57\\
\textsc{ESPRIT} & Ant. & Source &  Low & $2180{\rm E}-6^{\circ}$ & $2.70{\rm E}-2^{\circ}$ & $87.98{\rm E}6$ & $87.98{\rm E}6$ &1.23\\
\textsc{U-ESPRIT} &  &  & SNR & $1234{\rm E}-6^{\circ}$ & $7.31{\rm E}-2^{\circ}$ &$301.83{\rm E}6$ & $301.83{\rm E}6$ &1.16\\
\textsc{R-MUSIC} &  &  & & $6.63{\rm E}-6^{\circ}$ & $2.13{\rm E}-2^{\circ}$ & $127.45{\rm E}6$ & $135.37{\rm E}6$ & 1.75\\
\textsc{FT-DOA} &  &  & & $278.35{\rm E}-6^{\circ}$ & $5.15{\rm E}-2^{\circ}$ & $0.215{\rm E}6$ & $15.82{\rm E}6$ & 4.32\\
\hline
\hline
\textsc{DS} &  &  & & $472{\rm E}-12^{\circ}$ & $4.02{\rm E}-1^{\circ}$ & $66.30{\rm E}6$ & $157.82{\rm E}6$ & 1.50\\
\textsc{MVDR} &  & & (5dB) & $9.52{\rm E}-9^{\circ}$ & $4.54{\rm E}-1^{\circ}$ & $66.63{\rm E}6$ & $158.09{\rm E}6$ &1.62\\
\textsc{MUSIC} &256 &1 & & $0.75{\rm E}-12^{\circ}$ & $6.60{\rm E}-3^{\circ}$ & $110.47{\rm E}6$ & $194.11{\rm E}6$ & 1.57\\
\textsc{ESPRIT} & Ant.& Source & High & $878{\rm E}-9^{\circ}$ & $6.70{\rm E}-4^{\circ}$ & $87.98{\rm E}6$ & $87.98{\rm E}6$ & 1.23\\
\textsc{U-ESPRIT} &  &  & SNR & $899{\rm E}-9^{\circ}$ & $7.17{\rm E}-4^{\circ}$ &$301.83{\rm E}6$ & $301.83{\rm E}6$ & 1.16\\
\textsc{R-MUSIC} &  & & & $16{\rm E}-9^{\circ}$ & $1.39{\rm E}-4^{\circ}$ & $127.45{\rm E}6$ & $135.37{\rm E}6$ & 1.75\\
\textsc{FT-DOA} &  &  & & $55.6{\rm E}-6^{\circ}$ & $5.23{\rm E}-3^{\circ}$ & $0.215{\rm E}6$ & $15.82{\rm E}6$ & 4.32\\
\hline
\end{tabular}}
\end{center}
}
\label{tab:resumo_resul}
\end{table}

An important metric for determining DoA discrimination (accuracy) has been established and applied to the seven DoA methods analyzed. This metric has enabled a fair comparison, even if they operate differently from each another. Hence, in terms of DoA discrimination, the R-MUSIC presents better results, although the {ESPRIT} performance is very close, while the complexities of MUSIC and U-ESPRIT are much higher, especially {MUSIC} when there are several sources to be discriminated and the FT-DoA complexity is much smaller than the others. Moreover, the ESPRIT method has attained an intermediate DoA discrimination {in high SNR} with complexity very close to that of Root-MUSIC.  Moreover, in terms of MSE figure-of-merit, the best accuracy results have been obtained with MUSIC, Root-MUSIC and DS.  

Indeed, such high-resolution MUSIC, ESPRIT, U-ESPRIT, R-MUSIC and FT-DoA methods have offered improved performance-complexity tradeoff with high SNR; whereas DS and MVDR methods do not attain significant improvement in terms of DoA discrimination under either low or high SNR. {Besides, about {\it Speedup} figure-of-merit U-ESPRIT and ESPRIT have the worst performance, DS, MVDR, MUSIC, and R-MUSIC have intermediate performance, and the FT-DOA demonstrates the best performance, achieving approximately three times more processing faster than other methods, with 24 processors in parallel for example.} 

In this way, it can be observed that the better performance-complexity tradeoffs under large number of antennas scenarios are attained by the ESPRIT, Root-MUSIC and FT-DoA methods, with the {\it best performance-complexity tradeoff} achieved by the FT-DoA method due to its very low complexity, relative reliable MSE performance {and high parallelism level in the computation of the algorithm.} Hence, such method becomes very convenient for DoA estimation applications in massive MIMO systems, {\it i.e.}, a system with hundreds or even few thousand of antennas.

\section*{Acknowledgements}
This work was supported in part by the National Council for Scientific and Technological Development (CNPq) of Brazil under Grants 304066/2015-0, and in part by CAPES - CoordenaÃ§Ã£o de AperfeiÃ§oamento de Pessoal de NÃ­vel Superior, Brazil (scholarship),  and by the Londrina State University - ParanÃ¡ State Government (UEL).


\begin{thebibliography}{10}

\bibitem{GODARA2004}
Godara, L.C.:
\newblock {Smart Antennas}. 1 edn.
\newblock CRC Press, Boca Raton-FL (2004)

\bibitem{monzingo}
Monzingo, R.A., Haupt, R.L., Miller, T.W.:
\newblock {Introduction to Adaptive Arrays}.
\newblock Scitech, Raleigh - NC (2011)

\bibitem{narrowband}
Fourtz, J., Spanias, A., Banavar, M.K.:
\newblock {Narrowband Direction of Arrival Estimation for Antenna Arrays}. 1
  edn.
\newblock Morgan {\&} Claypool, Arizona (2008)

\bibitem{capon69}
Capon, J.:
\newblock {High-Resolution Frequency-Wavenumber Spectrum Analysis}.
\newblock Proceedings of the IEEE \textbf{57}(8) (1969)  1408--1418

\bibitem{reddy87}
Reddy, V.U., Paulraj, A., Kailath, T.:
\newblock {Performance Analysis of The Optimum Beamformer in The Presence of
  Correlated Sources and Its Behavior Under Spatial Smoothing}.
\newblock IEEE Transactions on Acoustics, Speech, and Signal Processing
  \textbf{35}(7) (1987)  927--936

\bibitem{Schmidt1986}
Schmidt, R.O.:
\newblock {Multiple emitter location and signal parameter estimation}.
\newblock IEEE Transactions on Antennas and Propagation \textbf{34}(3) (1986)
  276--280

\bibitem{godara}
Godara, L.C.:
\newblock {Application of antenna arrays to mobile communications, part II:
  Beam-forming and direction-of-arrival considerations}.
\newblock Proceedings of the IEEE \textbf{85}(8) (1997)  1195--1245

\bibitem{haykin}
Haykin, S.:
\newblock {Adaptative Filter Theory}. 3 edn.
\newblock Prentice Hall, New York, NY (1996)

\bibitem{rappaport}
Liberti, J.C., Rappaport, T.S.:
\newblock {Smart Antennas for Wireless Communications - IS-95 and Third
  Generation CDMA Applications}. 1 edn.
\newblock Prentice Hall, Upper Saddle River - NJ (1999)

\bibitem{FGross2015}
{Frank B. Gross}:
\newblock {Smart Antenna with MATLAB}. 2 edn.
\newblock Mc Graw Hill, New York, NY (2015)

\bibitem{Barabell1983}
Barabell, A.J.:
\newblock {Improving the resolution performance of eigenstructure-based
  direction-finding algorithms}.
\newblock ICASSP '83. IEEE International Conference on Acoustics, Speech, and
  Signal Processing \textbf{8} (1983)  8--11

\bibitem{Roy1989}
Roy, R., Kailath, T.:
\newblock {ESPRIT - Estimation of Signal Parameters Via Rotational Invariance
  Techniques}.
\newblock IEEE Transactions on Acoustics, Speech, and Signal Processing
  \textbf{37}(7) (1989)  984--995

\bibitem{Haardt1997}
Haardt, M.:
\newblock {Efficient One-, Two-, and Multidimensional High-Resolution Array
  Signal Processing}. 1 edn.
\newblock Shaker Verlag, Munich (1997)

\bibitem{Balanis2007}
Balanis, C.A., Ioannides, P.I.:
\newblock {Introduction to Smart Antennas}. 1 edn.
\newblock Morgan {\&} Claypool, Arizona (2007)

\bibitem{BenAllen2006}
Allen, B., Ghavami, M.:
\newblock {Adaptive Array Systems: Fundamentals and Applications}.
\newblock John Wiley {\&} Sons, West Sussex (2005)

\bibitem{two_stage2010}
Liang, J., Liu, D.:
\newblock {Passive Localization of Mixed Near-Field and Far-Field Sources Using
  Two-stage MUSIC Algorithm}.
\newblock IEEE Transactions on Signal Processing \textbf{58}(1) (jan 2010)
  108--120

\bibitem{two_stage2014}
Liu, G., Sun, X.:
\newblock {Two-Stage Matrix Differencing Algorithm for Mixed Far-Field and
  Near-Field Sources Classification and Localization}.
\newblock IEEE Sensors Journal \textbf{14}(6) (2014)  1957--1965

\bibitem{nested2010}
Pal, P., Vaidyanathan, P.P.:
\newblock {Nested arrays: A novel approach to array processing with enhanced
  degrees of freedom}.
\newblock IEEE Transactions on Signal Processing \textbf{58}(8) (2010)
  4167--4181

\bibitem{cadis2014}
Zhang, Y.D., Qin, S., Amin, M.G.:
\newblock {Doa estimation exploiting coprime arrays with sparse sensor
  spacing}.
\newblock In: ICASSP, IEEE International Conference on Acoustics, Speech and
  Signal Processing - Proceedings. Number~1, Florence, IEEE (2014)  2267--2271

\bibitem{unfolded_coprime2017}
Li, J., Zhang, X.:
\newblock {Direction of arrival estimation of quasi-stationary signals using
  unfolded coprime array}.
\newblock IEEE Access \textbf{5} (2017)  1--1

\bibitem{mailloux2005}
Mailloux, R.J.:
\newblock {Phased Array Antenna Handbook}. 2 edn. Volume~1.
\newblock Artech House, Norwood-MA (2005)

\bibitem{lcdoa_mm2016}
Yang, X., Liu, L., Wang, Y.:
\newblock {A new low complexity DOA estimation algorithm for massive MIMO
  systems}.
\newblock In: 2016 IEEE International Conference on Consumer Electronics-China
  (ICCE-China), Guangzhou, IEEE (dec 2016)  1--4

\bibitem{lcdoa_mm2016_2}
Meng, H., Zheng, Z., Yang, Y., Liu, K., Ge, Y.:
\newblock {A low-complexity 2-D DOA estimation algorithm for massive MIMO
  systems}.
\newblock In: 2016 IEEE/CIC International Conference on Communications in China
  (ICCC), Chengdu, IEEE (jul 2016)  1--5

\bibitem{lcdoa2016}
Ferreira, T.N., Netto, S.L., de~Campos, M.L., Diniz, P.S.:
\newblock {Low-complexity DoA estimation based on Hermitian EVDs}.
\newblock Proceedings of the IEEE Sensor Array and Multichannel Signal
  Processing Workshop (2016)  1--5

\bibitem{Zhang2011b}
Zhang, K., Ma, P., Zhang, J.Y.:
\newblock {DOA estimation algorithm based on FFT in switch antenna array}.
\newblock Proceedings of 2011 IEEE CIE International Conference on Radar
  \textbf{2}(4) (2011)  1425--1428

\bibitem{trench1989}
Trench, W.F.:
\newblock {Numerical Solution of the eigenvalue problem for hermitian toeplitz
  matrices}.
\newblock SIAM Journal Matrix Analisys Applications \textbf{10}(2) (1989)
  135--146

\bibitem{complex2007}
Hunger, R.:
\newblock {Floating Point Operations in Matrix-Vector Calculus}.
\newblock Technical report, Munich (2007)

\bibitem{cybenko1980}
Cybenko, G.:
\newblock {The Numerical Stability of The Levinson-Durbin Algorithm for
  Toeplitz Systems of Equations}.
\newblock SIAM Journal on Scientific Computing \textbf{1}(3) (1980)  303--319

\bibitem{haardt2014}
Haardt, M., Pesavento, M., Roemer, F., {Nabil El Korso}, M.:
\newblock {Subspace Methods and Exploitation of Special Array Structures}.
\newblock In: Academic Press Library in Signal Processing: volume 3; Array and
  Statistical Signal Processing.
\newblock 3 edn. Academic Press, Oxford (2014)  651--717

\bibitem{Amdahl1967}
Amdahl, G.M.:
\newblock {Validity of the single processor approach to achieving large scale
  computing capabilities}.
\newblock Proceedings of the April 18-20, 1967, spring joint computer
  conference on - AFIPS '67 (Spring) (1967)  483

\bibitem{Culler1997}
Culler, D., Singh, J.P., Gupta, A.:
\newblock {Parallel Computer Architecture: A Hardware/Software Approach}.
  Volume~1.
\newblock Morgan Kaufmann, San Francisco - CA, USA (1997)

\bibitem{Golub2013}
Golub, G.H., Loan, C.F.V.:
\newblock {Matrix Computations}. 4 edn.
\newblock Johns Hopkins University Press, Baltimore, MD (2013)

\end{thebibliography}
\end{document}